\documentclass[10pt, twocolumns]{IEEEtran}

\usepackage{amsfonts}
\usepackage{float}
\usepackage{color}
\usepackage{tikz}
\usepackage{graphicx}
\usepackage{epsfig}
\usepackage{amssymb}
\usepackage{amsmath}
\usepackage{multirow}
\usepackage{multicol}
\usepackage{cite}
\usepackage{bm}
\usepackage{color}
\usepackage{mathrsfs}
\usetikzlibrary{arrows}
\usepackage[ruled,vlined]{algorithm2e}

\newtheorem{proposition}{Proposition}

\newcommand{\ud}{\mathrm{d}}
\newcommand{\uD}{\mathrm{D}}
\newcommand{\scP}{\mathscr{P}}
\newcommand{\wtP}{{\widetilde{P}}}
\newcommand{\wtH}{{\widetilde{H}}}

\usepackage{booktabs}
\newcommand{\tabincell}[2]{\begin{tabular}{@{}#1@{}}#2\end{tabular}}

\begin{document}
\title{Mixed-ADC Massive MIMO Detectors: \\ Performance Analysis and Design Optimization}

\author{Ti-Cao Zhang, Chao-Kai Wen, Shi Jin,
    and~Tao~Jiang
    \thanks{
    	T.-C. Zhang and T.~Jiang are with the Wuhan National Laboratory for Optoelectronics, School of Electronic Information and Communications, Huazhong University of Science and
    	Technology, Wuhan, 430074, P. R. China. E-mail: ticao.hust@gmail.com; tao.jiang@ieee.org 
    	}
    \thanks{C.-K. Wen is with Institute of Communications Engineering, National Sun Yat-sen University, Taiwan. E-mail: chaokai.wen@mail.nsysu.edu.tw}
    \thanks{S. Jin is with the National Mobile Communications Research Laboratory, Southeast University, Nanjing 210096, China. E-mail: jinshi@seu.edu.cn}}
\maketitle


\begin{abstract}
Using a very low-resolution analog-to-digital convertor (ADC) unit at each antenna can remarkably reduce the hardware cost and power consumption of a massive multiple-input multiple-output (MIMO) system. However, such a pure low-resolution ADC architecture also complicates parameter estimation problems such as time/frequency synchronization and channel estimation. A mixed-ADC architecture, where most of the antennas are equipped with low-precision ADCs while a few antennas have full-precision ADCs, can solve these issues and actualize the potential of the pure low-resolution ADC architecture. In this paper, we present a unified framework to develop a family of detectors over the massive MIMO uplink system with the mixed-ADC receiver architecture by exploiting probabilistic Bayesian inference. As a basic setup, an optimal detector is developed to provide a minimum mean-squared-error (MMSE) estimate on data symbols. Considering the highly nonlinear steps involved in the quantization process, we also investigate the potential for complexity reduction on the optimal detector by postulating the common \emph{pseudo-quantization noise} (PQN) model. In particular, we provide asymptotic performance expressions including the MSE and bit error rate for the optimal and suboptimal MIMO detectors. The asymptotic performance expressions can be evaluated quickly and efficiently; thus, they are useful in system design optimization. We show that in the low signal-to-noise ratio (SNR) regime, the distortion caused by the PQN model can be ignored, whereas in the high-SNR regime, such distortion may cause 1-bit detection performance loss. The performance gap resulting from the PQN model can be narrowed by a small fraction of high-precision ADCs in the mixed-ADC architecture.
\end{abstract}


\IEEEpeerreviewmaketitle

\section{Introduction}\label{sec 1}

Massive multiple-input multiple-output (MIMO) systems are widely regarded as a disruptive technology for next-generation (i.e., 5G) communication systems \cite{Andrews-14JSAC,Wang-14COM-Mag,Sanguinetti-15JSAC}. By equipping a base station (BS) with an unprecedented number of antennas (a few hundreds or a thousand) in a centralized \cite{marzetta2010noncooperative,Hoydis-13JSAC} or distributed \cite{Zhang-13JSAC} fashion, such a system can reduce cell interference substantially through the simplest signal processing method because the channel vector between the users and the BS becomes quasi-orthogonal.

However, a large number of antennas significantly complicate the design of hardware for the implementation of massive MIMO in \emph{production}. In particular, such systems require an analog-to-digital converter (ADC) unit for each receiver antenna; therefore, using many antennas results in a need for an equivalent number of ADCs. The exponential increase in cost and power consumption attributed to high-speed and high-resolution ADCs\footnote{High-speed ADCs with a resolution of $\kappa$ bits typically adopt a flash architecture where the input voltage is compared with each of the $2^{{\kappa}}$ tap voltages simultaneously \cite{lee2008analog}.} is a major bottleneck in deploying massive MIMO systems \cite{Liu-15}. A solution to this problem involves using a very low-resolution ADC
(e.g., 1--3 bit) unit at each radio frequency chain \cite{risi2014massive,Bjornson-14TIT,Zhang-15TCOM,mezghani2007ultra,mo2014capacity,Jacobsson-ICC15}.

{\bf Previous work}: Low-resolution ADCs have favorable properties such as reduced circuit complexity, low power consumption, and feasible implementability. However, these converters inevitably deteriorate performance and complicate receiver design. The effect of low-resolution ADCs on channel capacity has been studied for single-input single-output (SISO) channels \cite{singh2009limits} and, recently, MIMO channels \cite{mezghani2007ultra,mo2014capacity,Jacobsson-ICC15}. Through a single antenna with 1-bit quantization at the receiver, channel capacity can be achieved by QPSK signaling \cite{singh2009limits}. In 1-bit MIMO cases, however, high-order constellations can be used to generate rates higher than those produced with QPSK signaling \cite{mo2014capacity,Jacobsson-ICC15}. In addition to information-theoretical studies, other works related to estimation/detection based on low-resolution samples include time/frequency synchronization \cite{Wadhwa-13Allerton}, channel estimation \cite{zeitler2012bayesian,Mo-14ACSSP,wen2015bayes,Choi-15Arxiv}, data detection \cite{mezghani2008maximum,Kamilov-12TSP,wang2013monobit,risi2014massive,Wang-15TWCOM,wen2015bayes,Choi-15Arxiv}, and related performance analysis \cite{mezghani2010belief,nakamura2008performance}.

The strongly nonlinear characteristic of the quantization process complicates the precise estimate of continuous variables. For example, a very long training sequence (requiring over 50 times the number of users) is necessary to achieve the same performance as the perfect CSI case in a MIMO system with 1-bit ADCs \cite{risi2014massive}. Such pilot overhead cannot be sustained by a practical system; to reduce pilot cost, a joint channel-and-data estimation method that used predicated payload data to aid channel estimation was proposed in \cite{wen2015bayes}. However, this technique enhances the computational complexity of the receiver. Other practical issues, such as automatic gain control (AGC) and time/frequency synchronization, have not been thoroughly examined in receivers with the pure low-resolution ADC architecture.

Motivated by the aforementioned considerations, we present a mixed-ADC receiver architecture named mixed-ADC massive MIMO system in the current study. In this architecture, most antennas were installed with low-resolution ADCs while a few antennas were equipped with full-resolution ADCs. A special case of the mixed-ADC massive MIMO system was initially proposed by \cite{liang2015mixed}; in this scenario, \emph{1-bit} ADCs replace the low-resolution ADCs. Under the mixed-ADC framework, CSI can be obtained in a round-robin manner \cite{liang2015mixed} in which high-resolution ADCs are connected to different antennas at various symbol times to estimate the corresponding channel coefficients. In the process, good-quality CSI is available at the receiver without significant pilots overhead. The available high-resolution chains can also assist in estimating other parameters and thus facilitate the establishment of several front-end designs, such as AGC and time/frequency synchronization. From an economic perspective, the mixed-ADC architecture is promising because it can be implemented by adding antennas with low-resolution ADCs to the existing BS.

{\bf Contributions}: In this study, we aim to examine mixed-ADC massive MIMO systems from a practical engineering perspective. In contrast to \cite{liang2015mixed}, which emphasizes the analysis of mutual information, our approach focuses on the MIMO (or multiuser) detection problem at the receiver. An extensive review of the large family of various MIMO detection algorithms is provided in \cite{yang2014fifty}, and MIMO detectors based on quantized samples are studied in \cite{Kamilov-12TSP,Wang-15TWCOM,mezghani2010belief,wen2015bayes,mezghani2008maximum,Choi-15Arxiv}. To the best of our knowledge, no research has shed light on MIMO detection problems in mixed-ADC architecture, the design and performance of which are in fact a central concern related to this architecture. This paper makes the
following specific contributions:
\begin{itemize}
    \item By exploiting probabilistic Bayesian inference, we provide a \emph{unified} framework to develop a family of MIMO detectors. This framework is labeled as the (generic) Bayes detector. To compute the Bayesian estimate, we must establish a prior distribution for the transmitted data and a likelihood function for the statistical model of receiver signals. Upon adopting the true prior and likelihood functions, the Bayes detector can achieve the best estimate in the mean squared error (MSE) sense. Properly postulating mismatched prior and likelihood functions can yield many low-complexity and popular detectors, such as the linear minimum mean-squared-error (LMMSE) and maximal-ratio-combining (MRC) receivers.

    \item The exact expression for the likelihood of a receiver signal with quantization is complex given the highly nonlinear property of the quantizer. The ``mixture'' architecture complicates the design of the Bayes detector further. A natural question is the following: \emph{how close to the best performance can a conventional MIMO detector operate without considering the exact (while annoying) nonlinear effect of the quantizers?} To answer this question, we adopt a traditional heuristic that treats quantization noise as additive and independent. This heuristic is known as the pseudo-quantization noise (PQN) model \cite[Chapter 4]{Widrow-BOOK}. By postulating a mismatched likelihood using this model in the Bayes detector, we can reduce computational cost significantly while degrading performance only slightly.

    \item To achieve the Bayes detector, we employ a recently developed technique called \emph{generalized approximate message passing} (GAMP) algorithm \cite{rangan2011generalized}. We adapt this approximation for the mixed-ADC architecture by specifying the corresponding adjustment in nonlinear steps. By applying the central-limit theorem (CLT) to the large system limit, we derive an approximate analytical expression for the \emph{state evolution} (SE) of the Bayes detector. A series of metrics, including bit error rate (BER) and MSE, can be predicted, and computer simulations are conducted to verify the accuracy of our analysis. The performance of the mixed-ADC massive MIMO receivers can be quickly and efficiently evaluated. Several useful observations are made based on the analysis to optimize the receiver design.
\end{itemize}


{\bf Notation}: Throughout this paper, vectors and matrices are presented in bold typeface, e.g., $\mathbf{x}$ and $\mathbf{X}$, respectively, while scalars are presented in regular typeface, e.g., $x$. We use $\mathbf{X}^T$ and $\mathbf{X}^*$ to respectively represent the transpose and conjugate transpose of a matrix $\mathbf{X}$. ${\rm Re}(\cdot)$ and ${\rm Im}(\cdot)$ respectively denote the real and imaginary parts of a complex matrix (vector). Normal distributions with mean $\mu$ and variance $\sigma^2$ are denoted by $\mathcal{N}(\mu,\sigma^2)$ while $\mathcal{CN}(\mu,\sigma^2)$ indicates a complex Gaussian distribution. Specifically, $\mathcal{N}(z;\mu,\sigma^2)$ [or $\mathcal{CN}(z;\mu,\sigma^2)$] denotes the probability density function (pdf) of a Gaussian random variable $z$ with mean $\mu$ and variance $\sigma^2$. Finally, let $\Phi(x)=\int_{-\infty}^x \uD t$ with $\uD t\triangleq \frac{1}{\sqrt{2\pi}}e^{-\frac{t^2}{2}}\ud t$.

\section{System Description}\label{sec 2}

\begin{figure}
    \begin{center}
        \includegraphics[width=3.5in]{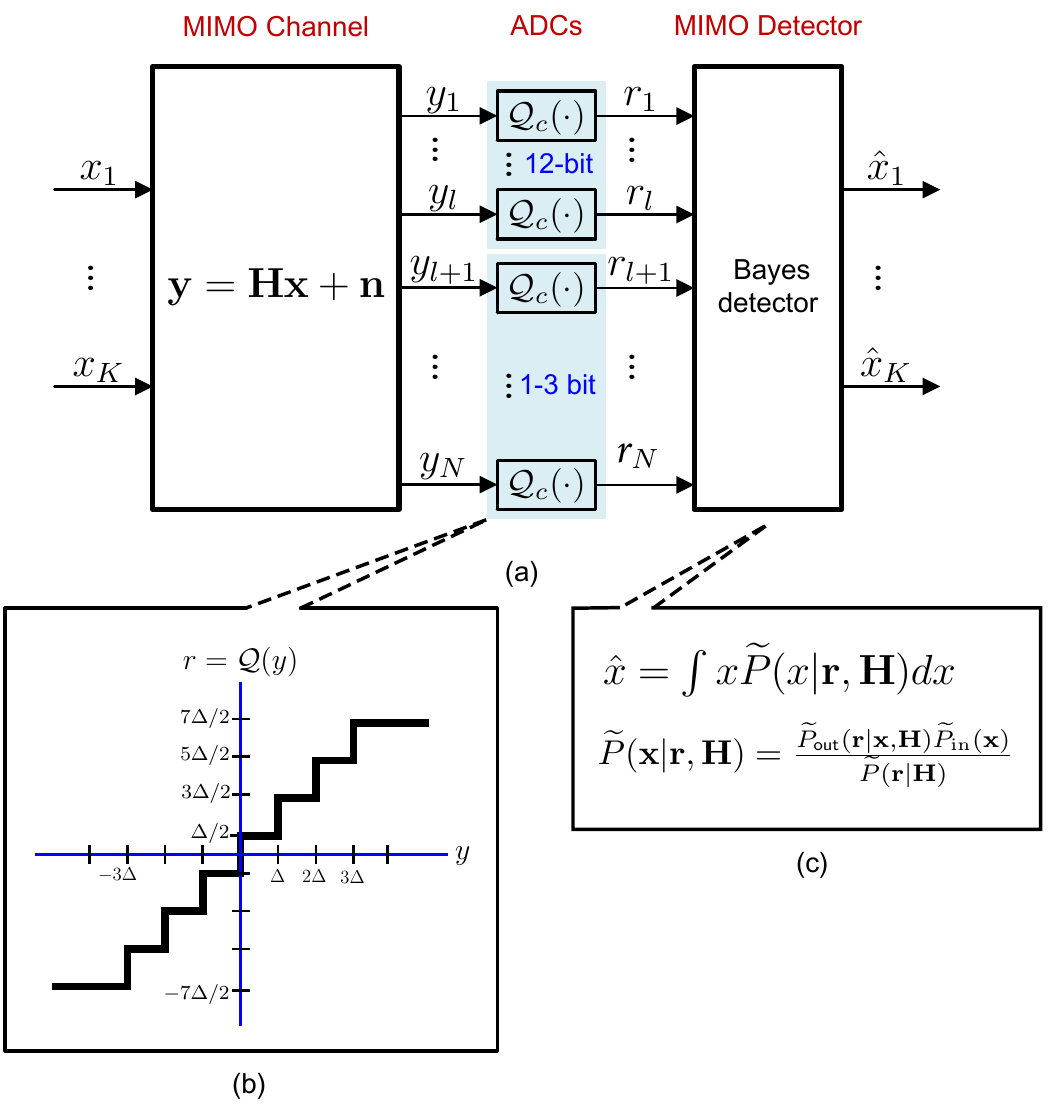}
        \caption{(a) A mixed-ADC massive MIMO architecture. (b) A 3-bit uniform  midrise quantizer. (c)  The (generic) Bayes detector. }\label{fig1}
    \end{center}
\end{figure}

We consider an uplink multiuser MIMO system that has $K$ single-antenna users and one BS equipped with an array of $N$ antennas. The discrete time complex baseband
received signal $\mathbf{y} \in \mathbb{C}^{N}$ is given by
\begin{equation}\label{2.1}
    \mathbf{y}=\frac{1}{\sqrt{K}}\mathbf{\wtH}\mathbf{x}+\mathbf{n}=\mathbf{H}\mathbf{x}+\mathbf{n},
\end{equation}
where ${\mathbf{x}=[x_j]\in\mathbb{C}^K}$ contains the transmitted symbols from all the users, ${\mathbf{n}=[n_i]\in\mathbb{C}^{N}}$ denotes the additive white Gaussian noise (AWGN), ${\mathbf{\wtH}=[\wtH_{ij}]\in\mathbb{C}^{N\times K}}$ represents the channel matrix between the BS and the $K$ users, and we further define $\mathbf{H} \triangleq \frac{1}{\sqrt{K}}\mathbf{\wtH}$\footnote{We take the variance of the channel coefficient to be $1/K$, to normalize the equivalent channel $\mathbf{H}=[H_{ij}]$ row and ensure that the analog baseband is within a proper range, e.g., ${(-1,\,+1)}$. This assumption is without loss of generality because in practice, a variable gain amplifier with automatic gain control is used before quantization. Moreover, the equivalent row normalized channel $\mathbf{H}$ simplifies the discussion in this study.}. The entries of $\mathbf{\wtH}$ are independent and identically distributed (i.i.d.) random variables with $\wtH_{ij}\sim\mathcal{CN}(0,1)$, the entries of $\mathbf{x}$ are i.i.d. and distributed as $P_{\sf in}(x)$, and the entries of $\mathbf{n}$ are i.i.d. $\mathcal{CN}(0,\sigma_n^2)$. If we suppose that $\mathbb{E}\{|x_j|^2\}=1, \,\forall j$, then the signal-to-noise ratio (SNR) of the system is defined by ${\rm SNR} = 1/\sigma_n^2$.

The real and imaginary components of the received signal at each antenna are quantized separately by an ADC. The quantization bits are set differently for
different antennas in the mixed-ADC massive MIMO system. All the $N$ complex valued quantizers are \emph{abstractly} described as $\mathcal{Q}_c(\cdot)$ so that
the quantized signal can be written as
\begin{equation}\label{2.3}
    \mathbf{r}=\mathcal{Q}_c(\mathbf{y}),
\end{equation}
where $\mathcal{Q}_c(\cdot)$ is applied element-wise and defined such that $r_i = \mathcal{Q}_c(y_i) \triangleq \mathcal{Q}({\rm Re}(y_i))+{\sf j}\mathcal{Q}({\rm
Im}(y_i))$. If a statement is for a generic antenna index, we often omit the subscript index $i$ from $r_i$ for brevity.

In this paper, we mainly focus on uniform midrise quantizers with quantization step size $\Delta$. Such a quantizer maps a real-valued\footnote{For ease of
notation, we abuse $r$ to denote each real channel although it should be specified as ${\rm Re}(r)$ or ${\rm Im}(r)$.} input that falls in $(r-\frac{\Delta}{2},
r+\frac{\Delta}{2}]$ to value $r$ from the discrete set
\begin{equation} \label{2.2}
    \mathcal{R}_{{\kappa}} \triangleq {\left\{ \Big({-\frac{1}{2}}+b\Big) \Delta; \,\, b=-\frac{2^{{\kappa}}}{2}+1, \cdots, \frac{2^{{\kappa}}}{2} \right\}} ,
\end{equation}
where $\kappa$ is the quantization bits. Figure~\ref{fig1}(b) shows a graphical depiction of a 3-bit uniform midrise quantizer. Notice that an input of magnitude
larger than ${(\frac{2^{\kappa}}{2}-1)}\Delta$ saturate. For ease of explication, we simply express the lower and upper thresholds associated with $r$ as $r^{\rm
low}$ and $r^{\rm up}$, respectively; specifically, they are
\begin{subequations}\label{quanbound}
    \begin{equation}
    r^{\rm low} = \left\{ \begin{array}{ll}
    r-\frac{\Delta}{2}, & \textrm{for $r\ge -{\left(\frac{2^{\kappa}}{2}-1\right)}\Delta$},\\
    -\infty, & \textrm{otherwise},
    \end{array} \right.
    \end{equation}
    and
    \begin{equation}
    r^{\rm up} = \left\{ \begin{array}{ll}
    r+\frac{\Delta}{2}, & \textrm{for $r\le {\left(\frac{2^{\kappa}}{2}-1\right)}\Delta$},\\
    -\infty, & \textrm{otherwise}.
    \end{array} \right.
    \end{equation}
\end{subequations}

We assume that most of the antennas adopt low-resolution ADCs (e.g., $1-3$ bit) while only a few antennas adopt high-resolution ADCs (e.g., $12$ bit). To
distinguish the various ADCs, we use $\Omega_{\kappa}$ to indicate the collection of the antennas equipped with ADCs of $\kappa$ bits. Moreover, the cardinality of
$\Omega_\kappa$ is $N_\kappa$, and thus we have $\sum_{\kappa}N_\kappa=N$.

Compared with the conventional massive MIMO system equipped with high-resolution ADCs, such a mixed-ADC architecture remarkably reduces the circuit cost and power
consumption. The available high-resolution chains can assist in estimating certain channel parameters. For example, CSI can be obtained in a round-robin manner
\cite{liang2015mixed} in which high-resolution ADCs are connected to different antennas at various symbol times to estimate the corresponding channel coefficients.
In the process, good-quality CSI is available at the receiver without significant pilots overhead. We assume that the realization of the channel is perfectly known by the BS; thus, we focus on the detection problem.

\section{MIMO Detectors}\label{sec 3}

To recover the multiuser signals $\mathbf{x}$ from the quantized measurement $\mathbf{r}$, we adopt the Bayesian inference. The Bayesian inference starts with
computing the posterior probability $\wtP(\mathbf{x}|\mathbf{r},\mathbf{H})$ according to the Bayes rule
\begin{equation}\label{2.4}
\wtP(\mathbf{x}|\mathbf{r},\mathbf{H})=\frac{\wtP_{\sf out}(\mathbf{r}|\mathbf{x},\mathbf{H})\wtP_{\rm in}(\mathbf{x})}{\wtP(\mathbf{r}|\mathbf{H})},
\end{equation}
where $\wtP(\mathbf{r}|\mathbf{H})=\int\ud \mathbf{x}\wtP_{\sf out}(\mathbf{r}|\mathbf{x},\mathbf{H})\wtP(\mathbf{x})$ is the marginal likelihood function,
$\wtP_{\sf out}(\mathbf{r}|\mathbf{x},\mathbf{H})$ is a likelihood function, and $\wtP_{\rm in}(\mathbf{x})$ is a measure of the input distribution $P_{\rm
in}(\mathbf{x})$. Here, $\wtP$ (including $\wtP_{\rm in}$ and $\wtP_{\sf out}$) indicates that the probability is different from the actual one, $P$.

Using the posterior probability, the Bayes estimate of the unknown vector $\mathbf{x}$ is the mean of the posterior distribution and its element is expressed as
\begin{equation}\label{2.5}
\hat{x}_j={\rm E}{\left\{x_j|\mathbf{r},\mathbf{H}\right\}},
\end{equation}
where the expectation over $x_j$ is w.r.t. the marginal posterior probability
\begin{equation}\label{2.6}
\wtP(x_j|\mathbf{r},\mathbf{H})=\int_{\mathbf{x}\setminus x_j}\ud \mathbf{x}\wtP(\mathbf{x}|\mathbf{r},\mathbf{H}).
\end{equation}
The notation ${\int_{\mathbf{x}\setminus x_j}\ud \mathbf{x}}$ denotes the integration over all the variables in $\mathbf{x}$ except for $x_j$.

Note that all our developed MIMO ``detectors'' are centred around the framework of (\ref{2.5}), which should be classified as an estimation problem rather than a
detection problem. However, transmitted signal $x_i$ is typically formed by a constellation modulation scheme such as quadrature amplitude modulation (QAM). In
such cases, the final constellation point can be determined from $\hat{x}_j$ via the maximum-likelihood decision rule. Therefore, with an abuse of terminology, we
call the framework for obtaining (\ref{2.5}) the (generic) Bayes ``detector''.

According to (\ref{2.4}), the posterior $\wtP(\mathbf{x}|\mathbf{r},\mathbf{H})$ is determined by $\wtP_{\sf out}(\mathbf{r}|\mathbf{x},\mathbf{H})$ and $\wtP_{\sf
in}(\mathbf{x})$. Postulating a mismatched measure $\wtP \neq P$ is due to a particular choice that can reduce computational complexity \cite{Guo-05TIT}. We
discuss these properties in the subsequent subsections. With reference the Bayes detector, we always mean (\ref{2.5}) with general $\wtP_{\sf out}$ and $\wtP_{\sf
in}$.

\subsection{De-Quantization (DQ)-optimal Detector}\label{sec 3.1}
We begin with the detector development from the true $P_{\sf out}$ and $P_{\rm in}$, i.e, no mismatch between $P$ and $\wtP$. Let $w_i =
\mathbf{h}_i^{*}\mathbf{x}$ with $\mathbf{h}_i^{*}$ be the $i$-th row of $\mathbf{H}$. According to (\ref{2.1}) and (\ref{2.3}), the likelihood function (or the
conditional probability distribution) of the quantized output given $\mathbf{x}$ is expressed as
\begin{align}\label{3.1}
P_{\sf out}(\mathbf{r}|\mathbf{x},\mathbf{H})
= \prod_{i=1}^{N}
& P_{\sf out}{\left({\rm Re}\{r_i\}|{\rm Re}\{w_i\}\right)}  \nonumber \\
& \times
P_{\sf out}{\left({\rm Im}\{r_i\}|{\rm Im}\{w_i\}\right)}.
\end{align}
where
\begin{equation} \label{3.2}
P_{\sf out}(r|w) = \frac{1}{\sqrt{\pi\sigma_n^2}}\int_{{r^{\rm low}}}^{r^{\rm up}}\ud y e^{-\frac{(y-w)^2}{\sigma_n^2}}.
\end{equation}
Using the true likelihood function (\ref{3.1}) in conjunction with the true prior distribution $P_{\rm in}$ to (\ref{2.4}), we obtain the exact posterior
probability $P(\mathbf{x}|\mathbf{r},\mathbf{H})$. In this case, $\hat{x}_j$ via (\ref{2.5}) can achieve the best estimates in terms of MSE. We refer to this
detector as the de-quantization (DQ)-optimal detector.

However, the DQ-optimal detector is not computationally tractable because the marginal posterior probability $P(x_j|\mathbf{r},\mathbf{H})$ in (\ref{2.6}) involves
a high-dimensional integral. The canonical solution to this problem is using \emph{belief propagation} (BP) by expressing the posterior probability
$P(\mathbf{x}|\mathbf{r},\mathbf{H})$ as a factor graph \cite{guo2008multiuser}. Unfortunately, the complexity of BP is still too high for practical application.
To solve this problem, BP was recently simplified as the GAMP algorithm \cite{rangan2011generalized} by using the second-order approximations at measurement nodes.
We resort to the GAMP algorithm as an iterative procedure to solve the marginal posterior probability in a recursive way. Algorithm \ref{alg:GAMP} provides a
high-level description of GAMP to perform the (generic) Bayes detector under the mixed-ADC architecture. Note that Algorithm \ref{alg:GAMP} can be used not only to
achieve the DQ-optimal detector but also to operate other detectors introduced subsequently.

\begin{algorithm}[!h]\label{alg:GAMP}  \footnotesize
  \caption{GAMP Algorithm}
  \SetKwInOut{Input}{input}
  \SetKwInOut{Output}{output}
  \SetKwInOut{Initialize}{initialize}
  \SetKwInOut{Definition}{definition}


  \Input{The received quantized signal, $\mathbf{r}$; \\
  The channel state information, $\mathbf{H}$;\\
  The prior distribution of the input signal, $\wtP_{\sf in}(x)$;\\
  The likelihood function, $\wtP_{\sf out}(r|w)$;}
  \BlankLine
  \Output{$\mathbf{x}^t$ }
  \BlankLine
  \Definition{$g(p,v_p) = \frac{\partial}{\partial p^{*}} \log{\left( \int \ud w \wtP_{\sf out}(r|w) e^{ - \frac{|w-p|^2}{v_p} } \right)}$;\\
  $g'(p,v_p) = -\frac{\partial}{\partial p } g(p,v_p)$;}
  \BlankLine
  \Initialize{$\mathbf{x}^0=\mathbf{0}$, $\mathbf{v}_x^{0}=\mathbf{1}$, $\mathbf{p}^{0}=\mathbf{0}$;
  }
  \BlankLine
  \For{$t = 1,\cdots, t_{\max}$ }{
  {Output Step}:\\
  \nl \hspace{0.15cm} $\mathbf{v}_p^t= |\mathbf{H}|^2 \mathbf{v}_x^{t-1}$\;
  \nl \hspace{0.15cm} $\mathbf{p}^t=\mathbf{H}\mathbf{x}^{t-1}-\mathbf{v}_p^t\circ g{\left(\mathbf{p}^{t-1},\mathbf{v}_p^t\right)}$\;
  \nl \hspace{0.15cm} $\mathbf{v}_z^t= g'{\left(\mathbf{p}^t,\mathbf{v}_p^t\right)}$\;
  \nl \hspace{0.15cm} $\mathbf{z}^t= g{\left(\mathbf{p}^t,\mathbf{v}_p^t\right)}$\;
  {Input Step}:\\
  \nl \hspace{0.15cm} $\mathbf{v}_s^t=\left[ (|\mathbf{H}|^{2})^T \mathbf{v}_z^t \right]^{-1}$\;
  \nl \hspace{0.15cm} $\mathbf{s}^t=\mathbf{x}^{t-1}+\mathbf{v}_s^t\circ (\mathbf{H}^{*} \mathbf{z}^t )$\;
  \nl \hspace{0.15cm} $\mathbf{v}_x^t={\rm Var} {\left\{ \mathbf{x} |\mathbf{s}^t,\mathbf{v}_s^t\right\}}$\;
  \nl \hspace{0.15cm} $\mathbf{x}^t={\rm E} {\left\{ \mathbf{x} |\mathbf{s}^t,\mathbf{v}_s^t\right\}}$\;
  }
\end{algorithm}

Here, $t$ and $t_{\max}$ represent the current iteration index and the maximum iteration times, respectively. We define ${|\mathbf{A}|^2 = [|A_{ij}|^2]}$, and
$\circ$ denotes the Hadamard product; i.e., $\mathbf{A}\circ \mathbf{B}=[A_{ij} B_{ij}]$. Note that $[\cdot]^{-1}$, $g(\cdot)$, $g'(\cdot)$, ${\rm E}\{\cdot\}$,
and ${\rm Var}\{\cdot\}$ are applied element-wise. To better understand the algorithm, we provide some intuition on each step of Algorithm \ref{alg:GAMP}. Lines
1--2 compute an estimate $\mathbf{p}$ of the product $\mathbf{Hx}$ and the corresponding variance $\mathbf{v}_p$. The first term of $\mathbf{p}$ is a plug-in
estimate of $\mathbf{Hx}$ and the second term provides a refinement by introducing Onsager correction in the context of AMP \cite{Krzakala-12JSM}. Using
$\{\mathbf{p},\mathbf{v}_p\}$, lines 3--4 then compute the posterior estimate of the residual ${\mathbf{y}-\mathbf{p}}$ and the inverse residual variances
$\mathbf{v}_z$, where $\mathbf{y}$ is the posterior estimate of the un-quantized received signal by considering the likelihood function $\wtP_{\sf out}$. Lines
5--6 then use these residual terms to compute $\mathbf{s}$ and $\mathbf{v}_s$, where $\mathbf{s}$ can be interpreted as an observation of $\mathbf{x}$ under an
AWGN channel with zero mean and variance of $\mathbf{v}_s$. Finally, lines 7--8 estimate the posterior mean $\mathbf{x}$ and variances $\mathbf{v}_x$ by
considering the prior $\wtP_{\sf in}$.

Using the likelihood function of (\ref{3.2}) to Algorithm \ref{alg:GAMP}, we can obtain analytic expressions of $g$ and $g'$, which are
\begin{subequations}\label{3.4}
    \begin{align}
    g(p,v_p)
    & = \frac{\widetilde{r}-p}{v_p+\sigma_n^2},\\
    g'(p,v_p)&=\frac{1}{v_p+\sigma_n^2}{\left(1-\frac{\widetilde{\sigma}^2}{v_p+\sigma_n^2}  \right)},
    \end{align}
\end{subequations}
where\footnote{Notice that $w$ is a complex-valued variable. The integral $\int_{r^{\rm low}}^{r^{\rm up}} \ud w$ in (\ref{eq:deffr}) is given by
\begin{equation*}
    \int_{{\rm Re}(r^{\rm low})}^{{\rm Re}(r^{\rm up})}
    \int_{{\rm Im}(r^{\rm low})}^{{\rm Im}(r^{\rm up})} \ud {\rm Re}(w) \ud {\rm Im}(w).
\end{equation*}
}
\begin{subequations} \label{eq:deffr}
    \begin{align}
\widetilde{r} &:= \frac{\int_{r^{\rm low}}^{r^{\rm up}}w\,\mathcal{CN}(w|p,\sigma_n^2+v_p)\ud w}{\int_{r^{\rm low}}^{r^{\rm up}}\mathcal{CN}(w|p,\sigma_n^2+v_p)\ud w}, \\
\widetilde{\sigma}^2 &:= \frac{\int_{r^{\rm low}}^{r^{\rm up}}{\left|w-\widetilde{r}\right|}^2\,\mathcal{CN}(w|p,\sigma_n^2+v_p)\ud w}{\int_{r^{\rm low}}^{r^{\rm up}}\mathcal{CN}(w|p,\sigma_n^2+v_p)\ud w}.
    \end{align}
\end{subequations}
Note that the integration interval $(r^{\rm low},\,r^{\rm up}]$ in (\ref{eq:deffr}) varies with different antennas because the quantization bits are set
differently for different antennas in the mixed-ADC architecture. The variance ${\rm Var}\{\cdot\}$ and expectation ${\rm E}\{\cdot\}$ in Algorithm \ref{alg:GAMP}
are performed with respect to
\begin{equation} \label{eq:Px}
P(x|s^t,v_s^t) = \frac{\mathcal{CN}(x|s^t,v_s^t)P_{\sf in}(x)}{ \int \ud x' \mathcal{CN}(x'|s^t,v_s^t)P_{\sf in}(x') }.
\end{equation}
If the QAM constellation with $M$ points is used, the prior is given by $P_{\sf in}(x) = 1/M$ for $x$ being the constellation points, and $\int \ud x'$ denotes the
integral w.r.t. the discrete measure.

Note that the calculations of $g(\cdot)$, $g'(\cdot)$, ${\rm E}\{\cdot\}$, and ${\rm Var}\{\cdot\}$ for the DQ-optimal detector are highly nonlinear. Explicit
expressions of the nonlinear functions are provided in \cite{wen2015bayes}. Although the DQ-optimal detector can achieve the best estimate in terms of MSE, the
highly nonlinear integration steps involved make the calculation difficult and in turn complicate the circuit. Moreover, in the mixed-ADC massive MIMO system,
different antennas adopt ADCs with different resolutions, which makes the quantization process even more complex. This inconvenience motivates us to import
mismatched measures in $P_{\sf out}$ and $P_{\sf in}$ to simplify the calculation.

\subsection{Pseudo-De-Quantization (PDQ)-optimal Detector}\label{sec 3.2}

The nonlinear calculations of $g(\cdot)$ and $g'(\cdot)$ are due to the integration in (\ref{3.2}). To simplify the calculations, one heuristic is to treat the
nonlinear quantization process as an additive quantization error; i.e.,
\begin{equation}\label{3.3}
\mathbf{r}= \mathcal{Q}_c(\mathbf{H}\mathbf{x}+\mathbf{n}) = \mathbf{H}\mathbf{x}+\mathbf{n}+\mathbf{q}
\end{equation}
with ${\mathbf{q}= [q_{i}] = [\mathcal{Q}_c(y_i)-y_i]}$ being the quantization error. This heuristic is known as the PQN model \cite{Widrow-BOOK}.

In the PQN model, the quantization error $q_i$ and its input $y_i$ are generally assumed to be independent, and $q_i$ is usually modeled as a complex Gaussian
random variable with zero mean and variance $\sigma_q^2$.\footnote{The variance of the quantization error on one side (real or imaginary) is usually modeled as
equal to that of a uniform distribution on a quantization interval $[-\frac{\Delta}{2},\,\frac{\Delta}{2}]$, i.e., $\sigma_q^2 = \frac{\Delta^2}{12}$.} As a
result, the measure $P_{\sf out}$ in (\ref{3.2}) can be written as
\begin{equation} \label{3.6}
\wtP_{\sf out}(r|w) = \frac{1}{\sqrt{\pi\gamma}} e^{-\frac{(r-w)^2}{\gamma}}
\end{equation}
with $\gamma=\sigma_q^2+\sigma_n^2$. By using $\wtP_{\sf out}$ of (\ref{3.6}) in Algorithm \ref{alg:GAMP}, we obtain
\begin{subequations} \label{eq:g_linear}
    \begin{align}
    g(p,v_p) &= \frac{r-p}{v_p+  \gamma }, \\
    g'(p,v_p) &= \frac{1}{v_p+ \gamma }.
    \end{align}
\end{subequations}
The highly nonlinear steps in (\ref{3.4}) now becomes simple linear steps (\ref{eq:g_linear}). Thus, the computational complexity is reduced significantly. We call
Algorithm \ref{alg:GAMP} with $g$ and $g'$ in (\ref{eq:g_linear}) the pseudo-de-quantization(PDQ)-optimal detector.

From the perspective of the algorithm itself, the PDQ-optimal detector is exactly the AMP algorithm \cite{donoho2009message} although adjusting $\gamma$ for each antenna is required to reflect different quantization bits for different antennas. Originally, the AMP algorithm was designed to recover a
signal from a linearly transformed and additive Gaussian noise corrupted measurement. The resulting PDQ-optimal detector is induced from the GAMP algorithm by
postulating the \emph{mismatched} measure in $P_{\sf out}$, which is mainly due to the low-complexity purpose.


\subsection{Linear Detector}\label{sec 3.3}
In the PDQ-optimal detector, $g$ and $g'$ are linearized by postulating a mismatched measure in $P_{\sf out}$ while the calculation of $\{ \mathbf{x}, \mathbf{v}_x
\}$ in lines 7--8 of Algorithm \ref{alg:GAMP} are still nonlinear. Next, we purpose a further complexity reduction on the nonlinear steps by imposing an additional
mismatch in $P_{\sf in}$. In fact, if we postulate that the input is Gaussian, i.e., $\wtP_{\sf in}(x)=\mathcal{CN}(x;0,1)$, then the Bayes detector over the PQN
model (\ref{3.3}) becomes a linear detector. That is, ${x}_j$ in (\ref{2.5}) becomes
\begin{equation}\label{3.7}
\hat{\mathbf{x}}=\left(\mathbf{H}^*\mathbf{H}+\gamma\mathbf{I}\right)^{-1}\mathbf{H}\mathbf{r},
\end{equation}
where the parameter $\gamma$ is commonly called the regularization factor. In particular, when $\gamma$ is set to $0$ and $\infty$, respectively, we immediately
obtain the zero-forcing detector and the MRC detector \cite{Guo-05TIT}.

The linear estimator (\ref{3.7}) can also be implemented via Algorithm \ref{alg:GAMP}. To this end, we substitute $\wtP_{\sf in}(x) \sim \mathcal{CN}(0,1)$ into
(\ref{eq:Px}), and then lines 7--8 of Algorithm \ref{alg:GAMP} become
\begin{equation} \label{eq:x_linear}
v^t_{x_j} = \frac{v^t_{s_j}}{1+v^t_{s_j}}, \qquad
x^t_j =\frac{s^t_j}{1+v^t_{s_j}}.
\end{equation}
Now, the nonlinear steps of ${\rm E}\{\cdot\}$ and ${\rm Var}\{\cdot\}$ become simple linear steps (\ref{eq:x_linear}).

The computational complexity order for the dimensional matrix inversion in (\ref{3.7}) is $O(NK^2)$. When the linear estimator is implemented using Algorithm
\ref{alg:GAMP}, the computational complexity order is $O(NKt_{\max})$. In fact, Algorithm \ref{alg:GAMP} converges very quickly after 8--10 iterations. Thus, the
computational complexity by using Algorithm \ref{alg:GAMP} is smaller than that of the conventional linear detector (\ref{3.7}) using the matrix inversion. In
addition, the iteration steps in Algorithm \ref{alg:GAMP} only involves matrix-vector multiplications that are highly suitable for hardware implementation.

\section{Performance Analysis}\label{sec 4}

\begin{table}\caption{Special Cases for the Generalized Bayes Estimator }
    \label{table 1}
    \begin{center}
        \begin{tabular} {lcccc}
            \toprule
            \multirow{2}{*}{\textbf{Method}}  & \multicolumn{2}{c}{$\wtP_{\sf in}$}  & \multicolumn{2}{c}{$\wtP_{\sf out}$} \\ \cmidrule(lr){2-3}\cmidrule(l){4-5}
            & distribution & mismatch & distribution & mismatch \\
            \midrule
            \textbf{DQ-Opt.}    & $P_{\sf in}(x)$  & No   & eq.(\ref{3.2})  & No\\
            \textbf{PDQ-Opt.}     & $P_{\sf in}(x)$  & No   & eq.(\ref{3.2})  & Yes\\
            \textbf{Linear}  & Gaussian         & Yes  & eq.(\ref{3.6})  & Yes\\
            \bottomrule
        \end{tabular}
    \end{center}
\end{table}

We have constructed numerous detectors that range from the optimal (but complex) detector to the suboptimal (but simple) MRC detector under a unified framework.
All of these detectors are induced from the Bayesian inference by a particular choice of $\wtP_{\sf in}$ and $\wtP_{\sf out}$, the features of which are summarized
in Table \ref{table 1}. Remarkably, these detectors can be realized by the GAMP algorithm (i.e., Algorithm \ref{alg:GAMP}); thus, their performance are amenable to
asymptotic analysis. Specifically, the GAMP dynamics can be tracked by a set of state evolution (SE) equations \cite{rangan2011generalized}.

In Section IV-A, we derive the SE equations for the \emph{generic} Bayes detector in the mixed-ADC massive MIMO system. In Section IV-B, we particularize the SE
equations for the three detectors constructed in Section III. Some performance comparisons among the three detectors are also provided in this subsection.

\subsection{State Evolution Analysis}\label{sec 4.1}

In this section, we derive the SE equations of Algorithm \ref{alg:GAMP} under the mixed-ADC massive MIMO system. Our derivation is performed in the large-system
regime where $K$ and $N$ reach infinity while the ratios
\begin{equation}
 {N}/{K} \rightarrow \lambda, \quad {N_{\kappa}}/{K} \rightarrow \lambda_{\kappa}, ~\forall \kappa
\end{equation}
remain fixed.

We perform the analysis starting from the algorithm itself. Recall that output $\mathbf{x}^{t}$ is deduced from line 7 of Algorithm \ref{alg:GAMP}. In lines 7--8
of Algorithm \ref{alg:GAMP}, the posterior averages are evaluated with respect to (\ref{eq:Px}), which is determined by $(s_j^{t},v_{s_j}^{t})$. Thus, we need to
determine the asymptotic behaviors for $s_j^{t}$ and its associated variance $v_{s_j}^{t}$ at iteration $t$. Injecting the quantities appearing in $s_j^{t}$ term
in line 6 of Algorithm \ref{alg:GAMP}, we notice that $s_j^{t}$ can be represented as a sum of $x_j$ and many other terms. Using the CLT, we can characterize
$s_j^{t}$ by its mean and variance. A proof is provided in Appendix \ref{Appendix A}. The result is summarized by the following proposition:

\begin{proposition} \label{Pro:pro1}
Given ${x_j = x \sim P_{\sf in}(x)}$, the asymptotic behavior of $s_j^{t}$ in Algorithm \ref{alg:GAMP} can be characterized by
\begin{equation}\label{4.1}
   \frac{D^t}{E^t}x+\frac{\sqrt{A^t}}{E^t}z,
\end{equation}
where $z\sim\mathcal{CN}(0,1)$. The parameters $A^t,D^t,E^t$ are independent of index $j$ and evolve as
\begin{subequations}\label{4.2} 
    \begin{align}
    & \hspace{-0.25cm} A^{t+1}=\sum_{\kappa}\lambda_{\kappa} {\left[\sum_{r\in \mathcal{R}_{\kappa}}
    \int\uD u \Psi{\left(r\left|\sqrt{\frac{|v_{x\hat{x}}^{t}|^2}{v_{\hat{x}}^{t}}}u\right.\right)} \Theta^2{\left(r\left|\sqrt{v_{\hat{x}}^{t}}u\right.\right)}  \right]},\label{4.2a} \\
    & \hspace{-0.25cm} D^{t+1}=\sum_{\kappa}\lambda_{\kappa} {\left[\sum_{r\in \mathcal{R}_{\kappa}}
    \int\uD u \Psi'{\left(r\Bigg|\sqrt{\frac{|v_{x\hat{x}}^{t}|^2}{v_{\hat{x}}^{t}}}u\right)} \Theta{\left(r\left|\sqrt{v_{\hat{x}}^{t}}u\right.\right)} \right]},\label{4.2b}\\
    & \hspace{-0.25cm} E^{t+1}=\sum_{\kappa}\lambda_{\kappa} {\left[\sum_{r\in \mathcal{R}_{\kappa}}
    \int\uD u \Psi{\left(r\Bigg|\sqrt{\frac{|v_{x\hat{x}}^{t}|^2}{v_{\hat{x}}^{t}}}u\right)} \Theta'{\left(r\left|\sqrt{v_{\hat{x}}^{t}}u\right.\right)} \right]},\label{4.2c}\\
    & v_x ={\rm E}_{x}{\left\{ |x|^2 \right\}},  \label{4.2d}  \\
    & c_{\hat{x}}^{t}={\rm E}_{x,z}{\left\{ f_2\left(\frac{D^{t}\cdot x+\sqrt{A^{t}}z}{E^{t}},\frac{1}{E^{t}}\right)\right\}}, \label{4.2e} \\
    &  v_{x\hat{x}}^{t}={\rm E}_{x,z}{\left\{x^{*} f_1\left(\frac{D^{t}\cdot x+\sqrt{A^{t}}z}{E^{t}},\frac{1}{E^{t}}\right)\right\}},\label{4.2f} \\
    & v_{\hat{x}}^{t}={\rm E}_{x,z}{\left\{ \left|f_1{\left(\frac{D^{t}\cdot x+\sqrt{A^{t}}z}{E^{t}},\frac{1}{E^{t}}\right)} \right|^2 \right\}}, \label{4.2g}
    \end{align}
\end{subequations}
where $\lambda_{\kappa}=N_{\kappa}/K$, $x \sim P_{\sf in}(x)$, $z\sim\mathcal{CN}(0,1)$, and
\begin{subequations}\label{4.3}
    \begin{align}
    &\Psi(r|\vartheta)=\Phi{\left(\frac{\sqrt{2}r^{\rm up}-\vartheta}{\sqrt{{\sigma_n^2+v_x-\frac{|v_{x\hat{x}}^{t}|^{2}}{v_{\hat{x}}^{t}}}}}\right)} -\Phi{\left(\frac{\sqrt{2}r^{\rm low}-\vartheta}{\sqrt{{\sigma_n^2+v_x-\frac{|v_{x\hat{x}}^{t}|^{2}}{v_{\hat{x}}^{t}}}}}\right)},  \label{4.3a}  \\
    &\Theta(r|\vartheta) = \frac{\partial }{\partial \vartheta} \log{\left(\int \ud w \wtP_{\sf out}(r|w) e^{ - \frac{(w-\vartheta)^2}{c_{\hat{x}}^{t}-v_{\hat{x}}^{t}} } \right)},\label{4.3b}
    \end{align}
\end{subequations}
with $\Psi'(r|\vartheta)=\frac{\partial \Psi(r|\vartheta)}{\partial \vartheta}$, $\Theta'(r|\vartheta)=-\frac{\partial \Theta(r|\vartheta)}{\partial \vartheta}$,
and
\begin{subequations}\label{4.4}
    \begin{align}
    & f_1(s,v)=\frac{\int \ud x x\wtP_{\sf in}(x)e^{-\frac{|x-s|^2}{v}}}{\int \ud x' \wtP_{\sf in}(x')e^{-\frac{|x'-s|^2}{v}}},\label{4.4a} \\
    & f_2(s,v)=\frac{\int \ud x |x|^2\wtP_{\sf in}(x)e^{-\frac{|x-s|^2}{v}}}{\int \ud x' \wtP_{\sf in}(x')e^{-\frac{|x'-s|^2}{v}}}.\label{4.4b}
    \end{align}
\end{subequations}
\end{proposition} \hspace{3.35in}\IEEEQEDopen

\begin{figure}
    \begin{center}
        \includegraphics[width=3.5in]{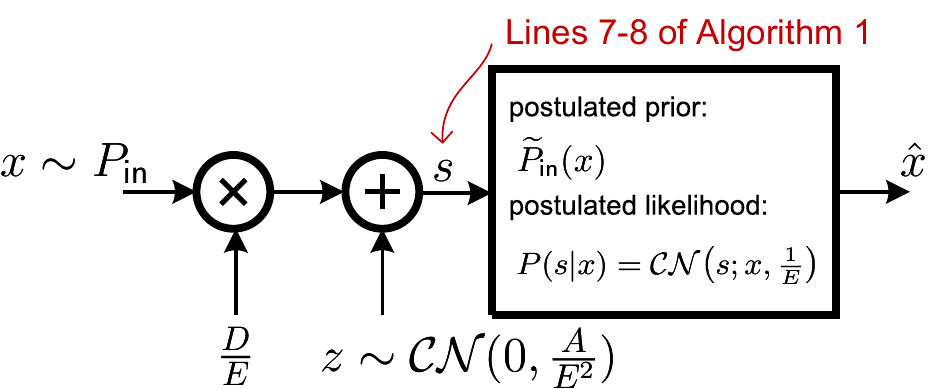}
        \caption{The equivalent channel sees at lines 7--8 of Algorithm \ref{alg:GAMP}.}\label{eqScalarChannel}
    \end{center}
\end{figure}

Before verifying the accuracy of the analysis, we closely examine Proposition \ref{Pro:pro1}, which has several valuable features.

(1) \emph{Analytical tractability}. Proposition \ref{Pro:pro1} reveals that the asymptotic behavior of $s_j^{t}$ in Algorithm \ref{alg:GAMP} can be characterized by the scalar equivalent model (\ref{4.1}). The scalar model (\ref{4.1}) is independent of index $j$. This characteristic indicates that lines 7--8 of Algorithm \ref{alg:GAMP} are decoupled into $K$ uncoupled scalar operators like the posterior average is over the scalar equivalent channel (\ref{4.1}) as shown in Figure~\ref{eqScalarChannel}. In the scalar equivalent channel, the Bayes detector postulates $\wtP_{\sf in}(x)$ and ${P(s|x) = {\cal CN}(s; x,\frac{1}{E})}$ as the prior distribution and the likelihood function, respectively. Note that the noise variance of the scalar channel is $\frac{A}{E^2}$ while the Bayes detector adopts $\frac{1}{E}$ as the noise variance because of the mismatched measure imported in $\wtP_{\sf out}$. Using the scalar equivalent channel, we can easily compute a large class performance metric for the Bayes detector such as MSE, BER, and mutual information. With regard to this performance metric, we are interested in the algorithm at time index $t_{\max} \rightarrow \infty$, and thus we omit the superscript index $t_{\max}$ from the related parameters for brevity.
We provide two examples as follows:
\begin{itemize}
    \item If $P_{\sf in}(x)$ is the QPSK signal, i.e.,
    \begin{equation}\label{inputdist}
    P_{\sf in}(x)=\frac{1}{4}\delta{\left(x\pm \frac{1}{\sqrt{2}}\pm{\sf j}\frac{1}{\sqrt{2}}\right)},
    \end{equation}
    then we can obtain the theoretical BER of the Bayes detector from (\ref{4.1}). That is,\footnote{For other high-order modulations, the closed-form of the
    BER expression can be derived based on the analysis in \cite{Cho-02TCOM}.}
    \begin{equation}\label{4.7}
    {\sf BER} =Q{\left(\frac{D}{\sqrt{A}}\right)},
    \end{equation}
    where $Q(z)=\frac{1}{\sqrt{2\pi}}\int_{z}^{+\infty}e^{-\frac{t^2}{2}}\ud t$ is the error function.
    \item For an arbitrary modulation signal, $f_1(\cdot)$ and $f_2(\cdot)$ in (\ref{4.2d})--(\ref{4.2g})
    are the conditional mean and second moment estimates of $x$ over the scalar channel (\ref{4.1}) while postulating mismatched prior and likelihood
    functions. As a result, the MSE of the Bayes detector can be shown as
    \begin{equation}\label{4.8}
    {\rm MSE}= {\rm E}{\left\{ |x-\hat{x}|^2\right\}} = v_x-2{\rm Re}\{ v_{x\hat{x}} \} + v_{\hat{x}}.
    \end{equation}
\end{itemize}

(2) \emph{Generality}. The analytical framework can apply to arbitrary postulated measures by substituting different prior distributions $\wtP_{\sf in}$ into (\ref{4.4}) and likelihood functions $\wtP_{\sf out}$ into (\ref{4.3b}). Moreover, $\Psi(r|x)$ given in (\ref{4.3a}) characterizes the \emph{true} quantization effect on an interval $(r^{\rm low},r^{\rm up}]$ and the summation $\sum_{r\in \mathcal{R}_\kappa}$ in (\ref{4.2a})-(\ref{4.2c}) adds all the corresponding interval together. The quantization bins and the quantization level are configurable. Thus, the SE equations can incorporate arbitrary quantization processes, such as nonuniform quantization. Moreover, $\sum_{\kappa}\lambda_{\kappa}$ in (\ref{4.2a})-(\ref{4.2c}) indicates the effect of the \emph{mixed}-ADC architecture. Interestingly, when particularizing our results to the case with a non-mixed ADC architecture, we recover the same asymptotic BER expression (\ref{4.7}) as found in \cite{nakamura2008performance} with the replica method.

(3) \emph{Computational simplicity}. The analytical result is computationally simple because the corresponding parameters (i.e., $A^t,D^t,E^t,c_{\hat{x}}^{t},
v_{x\hat{x}}^{t}, v_{\hat{x}}^{t}$) can be obtained in an iterative way, with each iteration only involving scalar summations (\ref{4.2a})--(\ref{4.2c}) and scalar
estimation computations (\ref{4.2d})--(\ref{4.2g}). In fact, using (\ref{4.1}), we can predict the SE in time of Algorithm \ref{alg:GAMP} in the mixed-ADC massive
MIMO system. Therefore, instead of performing  time-consuming Monte Carlo simulations to obtain the corresponding performance metrics, we can predict the
theoretical behavior by SE equations in a very short time.

\subsection{Explicit Expressions}\label{sec 4.2}
We now show special cases of the SE equations in Section \ref{sec 4.1}. To present a relatively intuitive analytical result, all the expressions and discussions in
this subsection are centered on a QPSK input distribution although our analysis can be incorporated with arbitrary input distributions.

\newsavebox\GAMPa
\begin{lrbox}{\GAMPa}
    \begin{minipage}{0.3\textwidth}
        \begin{align*}
        & c_{\hat{x}}=v_{x}=1 \\
        & v_{\hat{x}}=v_{x\hat{x}}=\int \uD u \tanh(\sqrt{A}u+D)
        \end{align*}
    \end{minipage}
\end{lrbox}

\newsavebox\GAMPb
\begin{lrbox}{\GAMPb}
    \begin{minipage}{0.3\textwidth}
        \begin{align*}
        & A=\sum_{\kappa}\lambda_{\kappa} \sum_{r\in \mathcal{R}_{\kappa}}\int \uD u
        \frac{{\left({ \Psi' ({r}|\sqrt{v_{\hat{x}}}u)}\right)}^2}{{\Psi ({r}|\sqrt{v_{\hat{x}}}u)}} \\
        & D=E=A
        \end{align*}
    \end{minipage}
\end{lrbox}

\newsavebox\AMPa
\begin{lrbox}{\AMPa}
    \begin{minipage}{0.2\textwidth}
        \begin{align*}
        v_x &=c_{\hat{x}}=1   \\
        v_{x\hat{x}} & =\int \uD u \tanh(\sqrt{A}u+D)\\
        v_{\hat{x}} & =\int \uD u \tanh^2(\sqrt{A}u+D)
        \end{align*}
    \end{minipage}
\end{lrbox}

\newsavebox\AMPb
\begin{lrbox}{\AMPb}
    \begin{minipage}{0.3\textwidth}
    $A$, $D$, and $E$ are given in (\ref{4.2a})--(\ref{4.2c})
    without the superscript iteration index $t$, and the terms $\Theta({y}|\vartheta)$ and $\Theta'({y}|\vartheta)$ can be simplified as
    \begin{subequations}\label{B.7}
        \begin{align}
            \Theta({y}|\vartheta)  & =\frac{y-\vartheta}{\gamma+c_{\hat{x}}-v_{\hat{x}}},\\
            \Theta'({y}|\vartheta) & =\frac{1}{\gamma+c_{\hat{x}}-v_{\hat{x}}}.
        \end{align}
    \end{subequations}
    \end{minipage}

\end{lrbox}

\newsavebox\Lineara
\begin{lrbox}{\Lineara}
    \begin{minipage}{0.2\textwidth}
        \begin{align*}
        v_x & = 1   \\
        c_x & = \frac{A+D^2}{(1+E)^2}+\frac{1}{1+E} \\
        v_{x\hat{x}} & = \frac{D}{1+E}\\
        v_{\hat{x}} & =\frac{A+D^2}{(1+E)^2}
        \end{align*}
    \end{minipage}
\end{lrbox}

\newsavebox\Note
\begin{lrbox}{\Note}
    \begin{minipage}{0.3\textwidth}
        $ \Psi(r|x)=\Phi{\left(\frac{\sqrt{2}r^{\rm up}-x}{\sigma_n^2+v_x-v_{\hat{x}}}\right)}
        -\Phi{\left(\frac{\sqrt{2}r^{\rm low}-x}{\sigma_{n}^2+v_x-v_{\hat{x}}}\right)}, $ and
        $ {\Theta({r}|x)}=e^{-\frac{(\sqrt{2}r-x)^2}{\gamma+c_{\hat{x}}-v_{\hat{x}}}}$
    \end{minipage}
\end{lrbox}

\begin{table*} \renewcommand{\arraystretch}{2.4}
    \centering
    \caption{Explicit Expressions for Generalized Bayes Estimators for QPSK constellation signals}
    \begin{center}
        \begin{tabular}{|l|l|l|}
            \hline
            \textbf{Method} & \multicolumn{2}{c|}{\textbf{SE parameters}}     \\
            \hline \hline
            \textbf{DQ-Opt.}   & \usebox{\GAMPa} & \usebox{\GAMPb} \\ \hline
            \textbf{PDQ-Opt.}    & \usebox{\AMPa} & \multirow{2}{*}{\usebox{\AMPb} } \\  \cline{1-2}
            \textbf{Linear} & \usebox{\Lineara}  &         \\ \hline
        \end{tabular}\label{table2}
    \end{center}
\end{table*}

For the DQ-optimal detector, the postulated measures are the same as the true measures for both $P_{\sf in}$ and  $P_{\sf out}$. Also, recall that the PDQ-optimal
detector only suffers a mismatched likelihood measure $\wtP_{\sf out}$ in (\ref{3.6}) and the linear detector further postulates the input as Gaussian.
Substituting the corresponding measures into Proposition \ref{Pro:pro1}, we obtain the explicit SE expressions for the DQ-optimal detector, the PDQ-optimal
detector, and the linear detector. The details of the expressions are listed in Table \ref{table2}. In particular, for the PDQ-optimal detector and the linear
detector, the term in the logarithm of (\ref{4.3b}) can be simplified as
\begin{equation}
    \int \ud w \wtP_{\sf out}(r|w) e^{ - \frac{(w-\vartheta)^2}{c_{\hat{x}}^{t}-v_{\hat{x}}^{t}} }
    =e^{-\frac{(\sqrt{2}r-\vartheta)^2}{\gamma+c_{\hat{x}}-v_{\hat{x}}}}.
\end{equation}
As a result, the terms $\Theta({y}|\vartheta)$ and $\Theta'({y}|\vartheta)$ in (\ref{4.2a})--(\ref{4.2c}) can be further simplified as (\ref{B.7}) shown in Table
\ref{table2}.

\begin{figure}[!t]
    \begin{center}
        \resizebox{3.50in}{!}{%
        	\includegraphics*{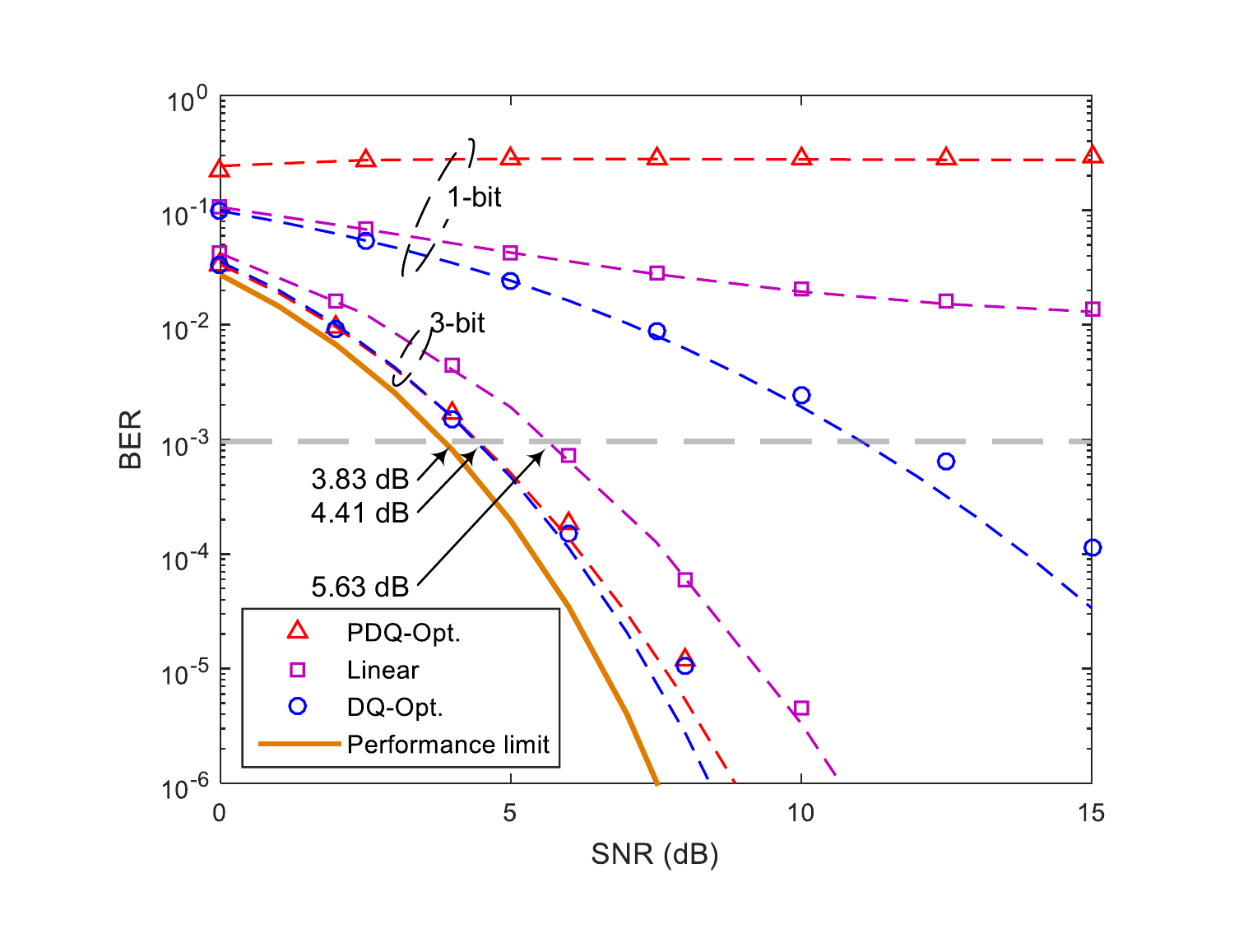} }%
        \caption{Comparison of analytical predictions against Monte Carlo simulations for various detectors. $N=200$, $K=50$, and $\Delta=0.5$ with a QPSK input. Dashed curve with color: theoretical prediction. Markers: simulation result. Solid curve: ultimate performance limit achieved by the DQ-optimal detector when the receiver has infinite precision to the received signal.}\label{fig3}
    \end{center}
\end{figure}

We now verify the accuracy of the analytical result presented in Table \ref{table2}. First, we evaluate the simplest non-mixed ADC case, i.e., the quantization function $\mathcal{Q}_{c}(\cdot)$ is the same for all antennas. In particular, we compare the BER expression (\ref{4.7}) with that obtained by computer simulations. The simulations are conducted over 10,000 channel realizations where Algorithm \ref{alg:GAMP} is performed with the maximum iteration times $t_{\max}=20$. We set the parameters as $N=200$ and $K=50$. In addition, we adopt a fixed step size $\Delta=0.5$.

As shown in Figure~\ref{fig3}, the BER results for various detectors constructed in Section \ref{sec 3} under different quantization levels are compared. The markers denote the numerical results while the dashed curve characterizes the analytical behaviors. The figure clearly shows precise predictions by our analytical expressions. Actually, we have conducted extensive simulation experiments to verify the accuracy of the analytical expressions with respect to the BER  (\ref{4.7}) and the MSE (\ref{4.8}) even including the mixed-ADC architecture. All of these examination plots are omitted because of space limitations.

Moreover, we observe that with the increase of quantization precision, all detectors generally achieve a significant performance improvement. For example, in the 1-bit quantization case, the PDQ-optimal detector and the linear estimators suffer from an error floor and almost do not work. However, in the 3-bit quantization case, the AMP estimator achieves an amazing performance gain; the gap between the DQ-optimal and PDQ-optimal detectors is extremely small. Even the linear detector only incurs a loss of $5.63-4.41=1.21$dB to that attained by the DQ-optimal detector when BER=$10^{-3}$. Recall that the PDQ-optimal detector drops out the integration parts in (\ref{3.2}) to reduce the computation cost. Figure~\ref{fig3} shows that the effect of this ``dropping" weakens with the increase of the resolution level. When the resolution goes to infinity, the integration effect is eventually eliminated, giving the PDQ-optimal and DQ-optimal detectors a common ultimate performance limit as predicted by the solid curve in Figure~\ref{fig3}. Therefore, instead of turning to the more sophisticated DQ-optimal detector, the PDQ-optimal detector with low precision (e.g., 3-bit) ADCs seems sufficient to achieve near optimal performance ($4.41-3.83=0.58$ dB loss at BER=$10^{-3}$ is acceptable) as the optimal \emph{unquantized} MIMO detector.

Finally, we point out an interesting result from Figure~\ref{fig3}. Theoretically, the linear detector should have poorer performance than the PDQ-optimal detector because the linear detector further imposes a mismatch in $P_{\sf in}$. Contrary to our belief, this inference seems to be un-true in the 1-bit quantization case. The reason, as shown in the following section, is that the PDQ-optimal detector is sensitive to the step size $\Delta$, which implies the importance of
parameter setting in the MIMO detectors.

\section{Discussions}\label{sec 5}
In this section, we discuss optimizing the MIMO detectors via the analytical expressions presented in Table \ref{table2}. The issues are related to quantization-step sizes, the regularization factor and the mixed receiver architecture.

\subsection{Quantization Step Size}\label{sec 5.1}
\begin{figure}[!t]
    \begin{center}
        \resizebox{3.50in}{!}{%
        	\includegraphics*{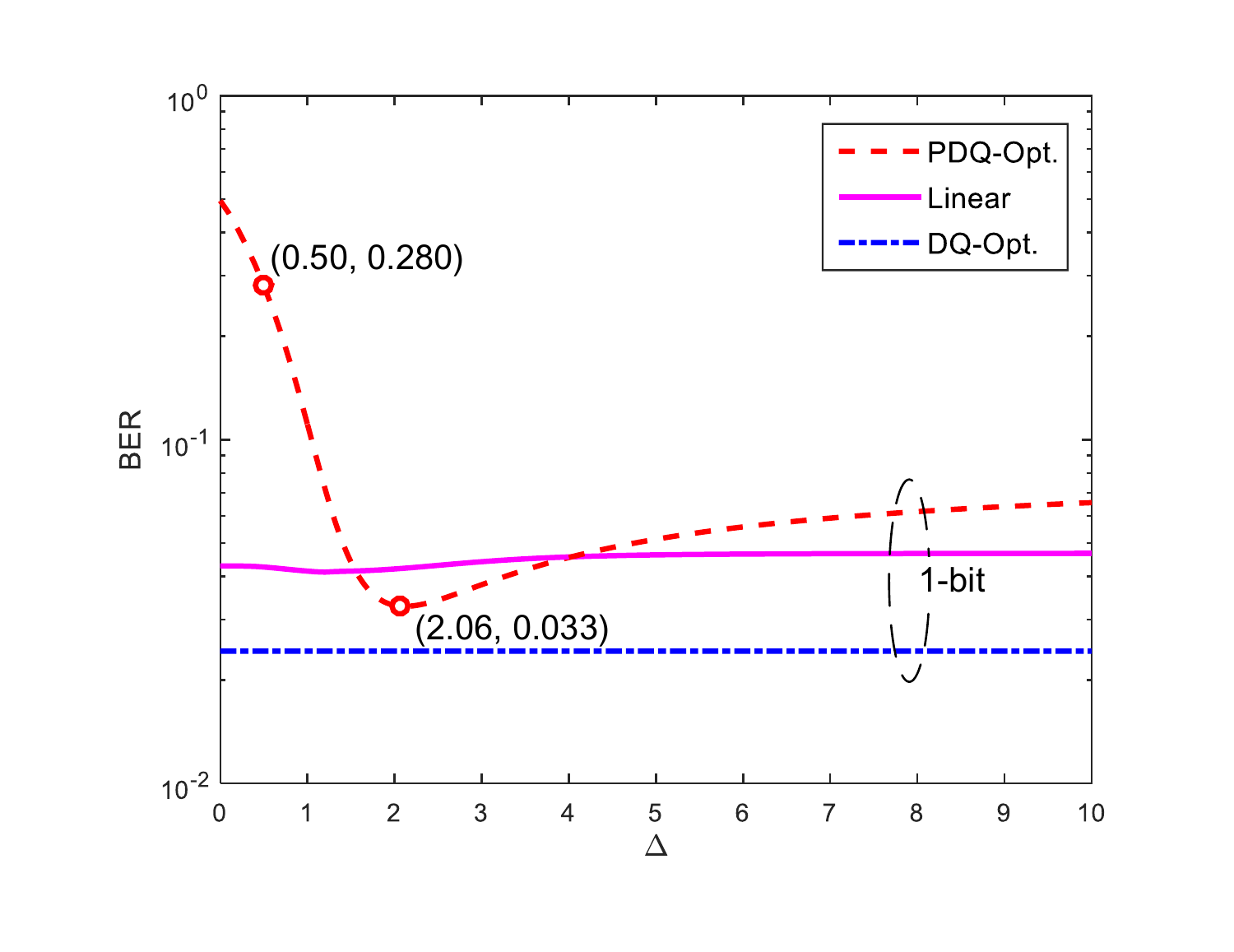} }%
        \caption{BER versus step size under 1-bit quantization with $\lambda = 4$, SNR = $5$ dB and a QPSK input.}\label{fig4}
    \end{center}
\end{figure}

We consider the 1-bit quantization case. Following a parameter setting similar to that shown in Figure~\ref{fig3}, Figure~\ref{fig4} depicts the average BER versus the step sizes under 1-bit quantization with SNR=5\,dB. The performance of the 1-bit DQ-optimal detector is \emph{irrelevant} to the value of step sizes. The reason is that the integrals in lines 3--4 in Algorithm \ref{alg:GAMP} [or, specifically, (\ref{3.4}) and (\ref{eq:deffr})] are taken w.r.t. either an interval $(-\infty,0]$ or $[0,+\infty)$ and only the sign of the output is of interest. However, this property is \emph{not} true for the other detectors. The performance of the PDQ-optimal detector and the linear detector are affected by the value of $\Delta$ although the former appears to be more sensitive. For example, when the step size is at the optimal value of 2.06, the PDQ-optimal detector achieves its best performance, which coincides with our inference that the performance of the PDQ-optimal detector is between that of the DQ-optimal detector and the linear detector. Recall that we choose the step size $\Delta=0.5$ in Figure~\ref{fig3} and the BER for the PDQ-optimal detector is 0.280. Thus, the PDQ-optimal detector does not work effectively in Figure~\ref{fig3}.

For the multi-bit uniform quantization, the optimum step sizes that minimize the corresponding BER for a QPSK input signal are compared to the input bit resolutions in Table \ref{tab:stepsize}. The corresponding step size values of the PDQ-optimal detector and the linear detectors are remarkably close for all bits because both detectors treat the quantization process as additive quantization error. This observation suggests that a prior mismatch may not cause a significant difference in terms of an optimized step size.

Recall that the element of a ADC input $y_{i}$ is the sum over many independent terms and thus can be approximated as a Gaussian distributed variable with variance $1+\sigma_n^2$. Therefore, we normalize the optimal step sizes according to \begin{equation}\label{normalizedstepsize}
\Delta_{\rm norm}=\frac{\sqrt{2}\Delta}{\sqrt{1+\sigma_n^2}}
\end{equation}
and average them over different SNRs. In this case, the step size is multiplied by $\sqrt{2}$ because the signal power is $\sqrt{2}$ times the power of the real or imaginary part. The corresponding normalization results are shown in Table \ref{table 3}. The optimal step sizes that try to minimize the quantization distortion for a Gaussian input signal \cite{al1996uniform} are listed in Table \ref{table 3} for comparison. The value of $\Delta_{\rm norm}$ is very close to those in \cite{al1996uniform} for all input bit rates $\kappa$ varying from 2 to 7. In fact, through extensive simulations under different antenna ratios, input constellations, and performance metrics (MSE or BER), we find that this property is general. Consequently, we can safely use the values in \cite{al1996uniform} as a consistent rule to assist the step size design. The rule of thumb does not apply to the 1-bit quantization case or the DQ-optimal detector. However, for these special cases, we can quickly obtain the optimal step sizes based on the analytical expressions.

\begin{table}\caption{Optimal step sizes for DQ-optimal\,/\,PDQ-optimal\,/\,linear detectors ($\lambda=4$).}\label{tab:stepsize}
    \centering
    \centering
    \begin{tabular}{rccc} \toprule
        \tabincell{c}{SNR\\(dB)} & {1 bit}  & {2 bit}   & {3 bit}  \\ \midrule \midrule
        -5   &   ---~~/\,4.247\,/\,5.150   &  1.990\,/\,1.461\,/\,1.438  &  1.147\,/\,0.849\,/\,0.845  \\
        -2.5 &   ---~~/\,3.245\,/\,3.595   &  1.611\,/\,1.217\,/\,1.173  &  0.932\,/\,0.697\,/\,0.689  \\
        0    &  ---~~/\,2.636\,/\,2.638    &  1.338\,/\,1.058\,/\,0.992  &  0.781\,/\,0.596\,/\,0.583  \\
        2.5  &   ---~~/\,2.271\,/\,1.925   &  1.132\,/\,0.954\,/\,0.874  &  0.674\,/\,0.531\,/\,0.514  \\ \midrule
        5    &  ---~~/\,2.057\,/\,1.519    &  0.970\,/\,0.886\,/\,0.801  &  0.594\,/\,0.489\,/\,0.475  \\
        7.5  &   ---~~/\,1.934\,/\,1.277   &  0.835\,/\,0.844\,/\,0.759  &  0.531\,/\,0.464\,/\,0.448  \\
        10   &  ---~~/\,1.865\,/\,1.131    &  0.717\,/\,0.818\,/\,0.735  &  0.474\,/\,0.449\,/\,0.432  \\
        12.5 &  ---~~/\,1.825\,/\,1.043    &  0.609\,/\,0.804\,/\,0.721  &  0.419\,/\,0.440\,/\,0.423  \\ \midrule
        15   & ---~~/\,1.803\,/\,0.990     &  0.511\,/\,0.795\,/\,0.713  &  0.366\,/\,0.436\,/\,0.419  \\
        17.5 &   ---~~/\,1.790\,/\,0.959   &  0.423\,/\,0.791\,/\,0.708  &  0.315\,/\,0.433\,/\,0.416  \\
        20   &  ---~~/\,1.783\,/\,0.942    &  0.347\,/\,0.788\,/\,0.706  &  0.267\,/\,0.431\,/\,0.415  \\ 
    \end{tabular}
    \\
    \centering
    \begin{tabular}{rccc}  \toprule
        \tabincell{c}{SNR\\(dB)} & {4 bit}  & {5 bit}  & {6 bit} \\ \midrule \midrule
        -5   &      0.651\,/\,0.484\,/\,0.483  &  0.365\,/\,0.272\,/\,0.271 &   0.202\,/\,0.150\,/\,0.150  \\
        -2.5 &      0.530\,/\,0.396\,/\,0.394  &  0.298\,/\,0.222\,/\,0.221 &   0.165\,/\,0.123\,/\,0.122  \\
        0    &     0.447\,/\,0.337\,/\,0.333  &  0.251\,/\,0.188\,/\,0.187 &   0.145\,/\,0.104\,/\,0.103  \\
        2.5  &      0.389\,/\,0.298\,/\,0.295  &  0.220\,/\,0.167\,/\,0.165 &   0.123\,/\,0.092\,/\,0.092  \\ \midrule
        5    &     0.349\,/\,0.274\,/\,0.272  &  0.199\,/\,0.153\,/\,0.152 &   0.111\,/\,0.085\,/\,0.084  \\
        7.5  &      0.318\,/\,0.260\,/\,0.256  &  0.184\,/\,0.145\,/\,0.145 &   0.104\,/\,0.080\,/\,0.079  \\
        10   &     0.292\,/\,0.251\,/\,0.247  &  0.171\,/\,0.145\,/\,0.145 &   0.098\,/\,0.077\,/\,0.077  \\
        12.5 &     0.267\,/\,0.246\,/\,0.245  &  0.160\,/\,0.145\,/\,0.145 &   0.092\,/\,0.076\,/\,0.075  \\ \midrule
        15   &    0.241\,/\,0.245\,/\,0.240  &  0.149\,/\,0.145\,/\,0.145 &   0.087\,/\,0.075\,/\,0.075  \\
        17.5 &      0.216\,/\,0.245\,/\,0.239  &  0.137\,/\,0.145\,/\,0.145 &   0.082\,/\,0.074\,/\,0.074  \\
        20   &     0.190\,/\,0.245\,/\,0.238  &  0.125\,/\,0.145\,/\,0.145 &   0.077\,/\,0.074\,/\,0.074  \\ \bottomrule
    \end{tabular}
\end{table}

\begin{table}
    \caption{Comparison the optimal step sizes for different purposes.}
    \begin{center}\label{table 3}
        \begin{tabular}{cccc} \toprule
            \tabincell{c}{$\kappa$ \\ (bits)} & \tabincell{c}{$\Delta_{\rm norm}$ \\ (PDQ-Opt.) }   & \tabincell{c}{$\Delta_{\rm norm  }$ \\ (Linear) }     & \tabincell{c}{$\Delta_{\rm opt.}$ \\ from \cite{al1996uniform} }         \\ \midrule \midrule
            2   &    1.0826  & 0.9619   &   1.0080   \\
            3   &    0.6014  & 0.5836   &   0.5895  \\
            4   &    0.3390  & 0.3345   &   0.3360  \\ \midrule
            5   &    0.1941  & 0.1936   &   0.1883 \\
            6   &    0.1042  & 0.1037   &   0.1041 \\
            7   &    0.0568  & 0.0568   &   0.0569 \\  \midrule
        \end{tabular}
    \end{center}
\end{table}

\subsection{Regularization Factor}\label{sec 5.2}

In the PQN model, the nonlinear quantization process is treated as additive quantization error with variance $\sigma_{q}^2$ in (\ref{3.6}) to simplify computation. So far, we have used the variance of a uniform distribution on a quantization interval $[-\frac{\Delta}{2},\,\frac{\Delta}{2}]$ to model $\sigma_{q}^2$, i.e.,
$\sigma_q^2=\frac{\Delta^2}{12}$. We now examine this approximation.

\begin{figure}[!t]
    \begin{center}
        \resizebox{3.50in}{!}{
        	\includegraphics*{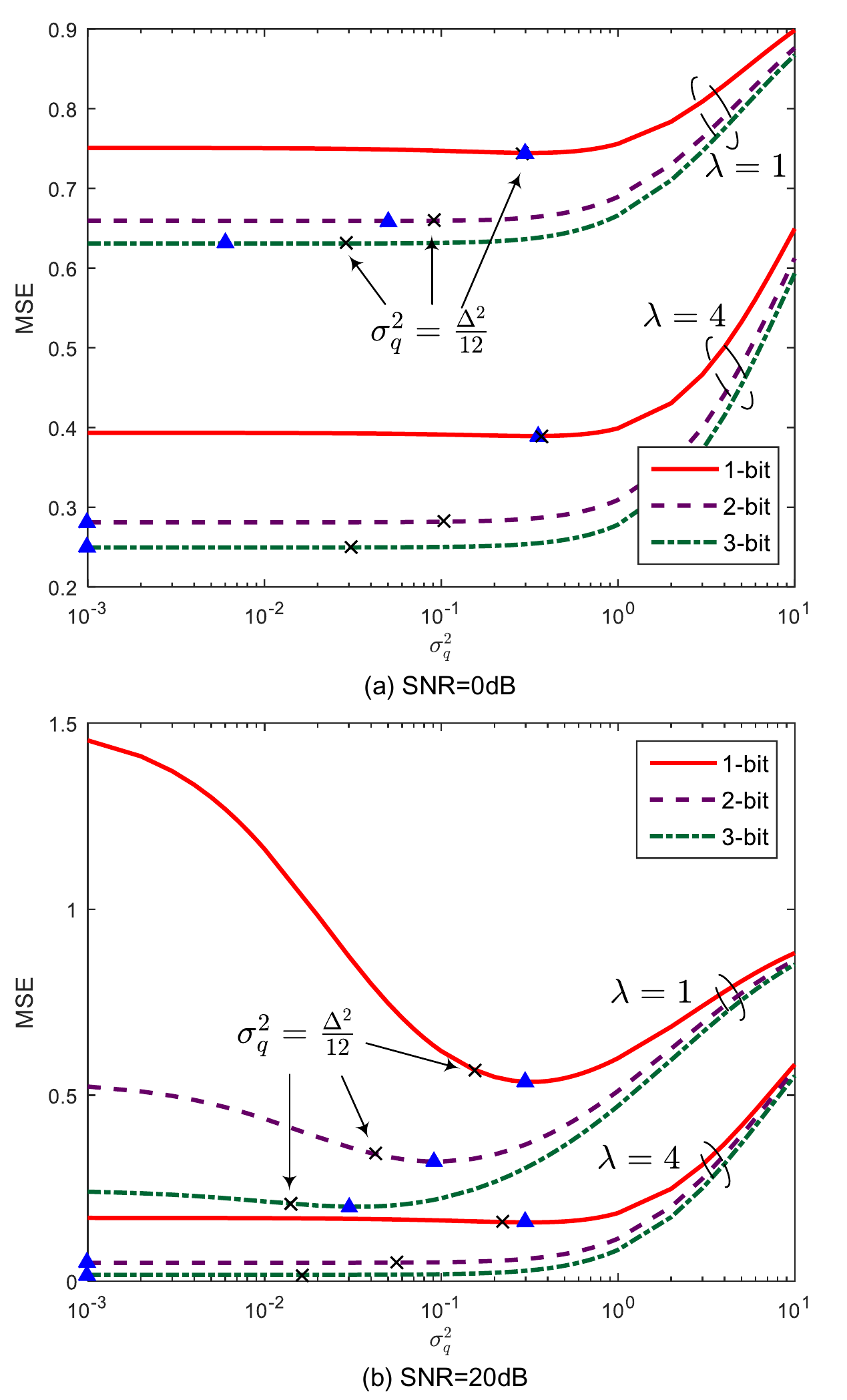} }%
        
        \caption{MSE of PDQ-optimal detector versus $\sigma_q^2$ under different quantization levels with a) SNR=$0$dB and b) SNR=$20$dB.
        The triangle markers correspond to the lowest MSE while the cross marks correspond to $\sigma_q^2 = \frac{\Delta^2}{12}$. }\label{fig5}
    \end{center}
\end{figure}

Figure~\ref{fig5} shows the MSEs of the PDQ-optimal detector with varying numbers of quantization bit depth $\kappa$ for Gaussian transmitted inputs.\footnote{In this case, the PDQ-optimal detector is exactly the linear detector because both the postulated and the actual inputs are Gaussian.} From the results in Figures \ref{fig5}(a) and \ref{fig5}(b), we find that in a typical massive MIMO system (e.g., $\lambda=4$), the MSEs of the PDQ-optimal detector are not sensitive to the value of $\sigma_q^2$ as long as its value is small enough, e.g., $\sigma_q^2<1$. However, for the case when the number of antennas are compatible with that of the users (e.g., $\lambda=1$), we should be more careful in selecting the value of $\sigma_q^2$. Comparing the triangle markers that correspond to the lowest MSE and the cross markers determined by $\sigma_q^2=\frac{\Delta^2}{12}$, we find that approximating $\sigma_q^2$ by the latter
rule is generally good enough. Although not precise, this rule of thumb seems effective.

\subsection{Mixed Architecture}\label{5.3}
We have discussed the optimal parameter settings for the quantization step size and the regularization factor. Using these optimal parameters, we now investigate the performance of the mixed-ADC architecture. Our discussions focus on the following question: \emph{how close to the best performance can a MIMO detector operate without considering the exact nonlinear effect of the quantizers?} In the following subsections, we refer to a system where most of the antennas adopt 1-bit resolution ADCs while the rest have full precision ADCs as \emph{the mixed 1-bit} architecture. We also have the \emph{mixed 2-bit} architecture and others.

\subsubsection{QPSK input}\label{5.3.1}
\begin{figure}[!t]
    \begin{center}
        \resizebox{3.50in}{!}{%
        	\includegraphics*{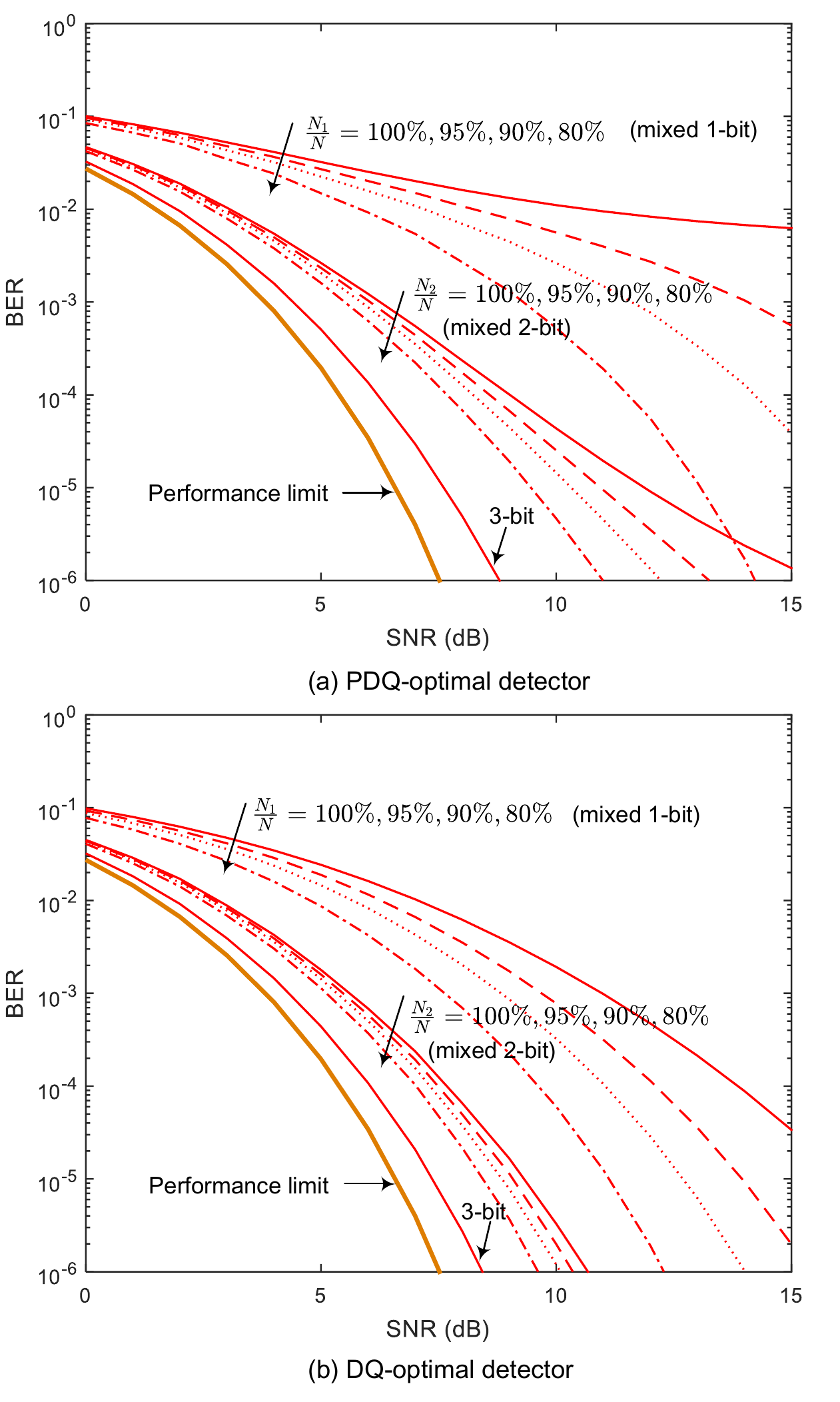} }
        \caption{BER versus SNR under a mixed-ADC architecture for different detectors with QPSK inputs.}\label{fig6}
    \end{center}
\end{figure}

Figure~\ref{fig6} illustrates the BER versus the SNR under different mixed architectures for QPSK inputs. Clearly, the mixed architecture helps improve the performance. For example, for the mixed 1-bit architecture, merely installing $5\%$ high-resolution ADCs (i.e., $\frac{N_1}{N}=95\%$) eliminates the error floor caused by the distortion of the mismatched measure of the PDQ-optimal detector as well as the 1-bit quantization. If we further increase the fraction of the full-precision ADCs to $10\%$ (i.e.,$\frac{N_1}{N}=90\%$), the mixed PDQ-optimal detector can achieve similar performance as the pure 1-bit DQ-optimal detector. Also, in the mixed 2-bit architecture, an $80\%$ 2-bit quantization mixed PDQ-optimal detector achieves a performance similar to the pure 2-bit DQ-optimal detector. Recall that the DQ-optimal detector is considered to achieve the best estimate in terms of minimizing the MSE. We conclude that the mixed architecture can help maintain the promised performance while significantly reducing the computational complexity by ignoring the exact but complex quantization process. In addition, comparing Figures \ref{fig6}(a) and \ref{fig6}(b), we notice that the mixed architecture is more advantageous to the PDQ-optimal detector than the DQ-optimal detector, which will also be validated by Figure~\ref{fig8}.


\subsubsection{Gaussian input}\label{5.3.2}
\begin{figure}[!t]
    \begin{center}
        \resizebox{3.50in}{!}{%
        	\includegraphics*{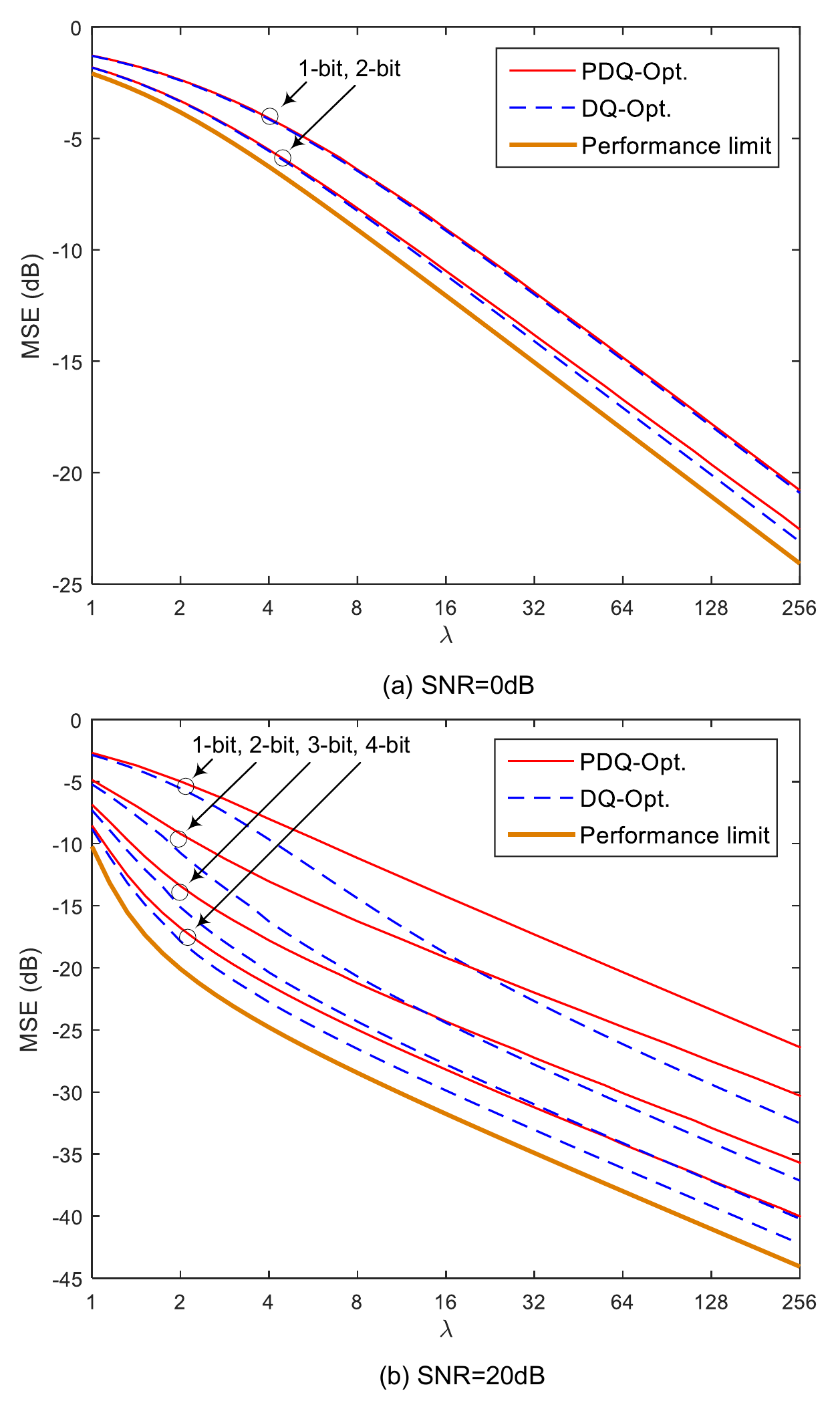} }%
        \caption{MSE for a Gaussian input signal versus MIMO configuration $\lambda=N/K$ with a) SNR=$0$\,dB and b) SNR=$20$\,dB.}\label{fig7}
    \end{center}
\end{figure}

In such an input type, the PDQ-optimal detector is exactly the linear detector because both the postulated and the actual inputs are Gaussian. We start by providing insights into the effect of quantization based on the non-mixed architecture. Figure~\ref{fig7} illustrates the MSEs of the detectors versus the antenna configuration ratios $\lambda =M/N$ for SNR=$0$\,dB and SNR=$20$dB. As shown in Figure~\ref{fig7}(a), when SNR=$0$\,dB, the performance degradation compared to the unquantized case caused by 1-bit quantization is approximately 3\,dB, indicating that ignoring the exact quantization effect is feasible in the low SNR regime. In the high SNR regime, however, a 1-bit resolution generally incurs significant performance losses as shown in Figure~\ref{fig7}(b). In a typical \emph{massive} MIMO system (e.g., $\lambda=16$), the PDQ-optimal detector generally causes 1-bit loss compared with the DQ-optimal detector. For example, the 2-bit PDQ-optimal detector has the same performance as the 1-bit DQ-optimal detector. Alternatively, the loss caused by the simplification of the quantization process can be compensated by doubling the number of receiver antennas. Figures \ref{fig7}(a) and \ref{fig7}(b) shows that the MSEs of the detectors generally improve by 3--6\,dB for each 1-bit rate increase. The higher the SNR is, the closer the MSE improvement is to 6\,dB.

\begin{figure}[!t]
    \begin{center}

        \resizebox{3.50in}{!}{%
        	\includegraphics*{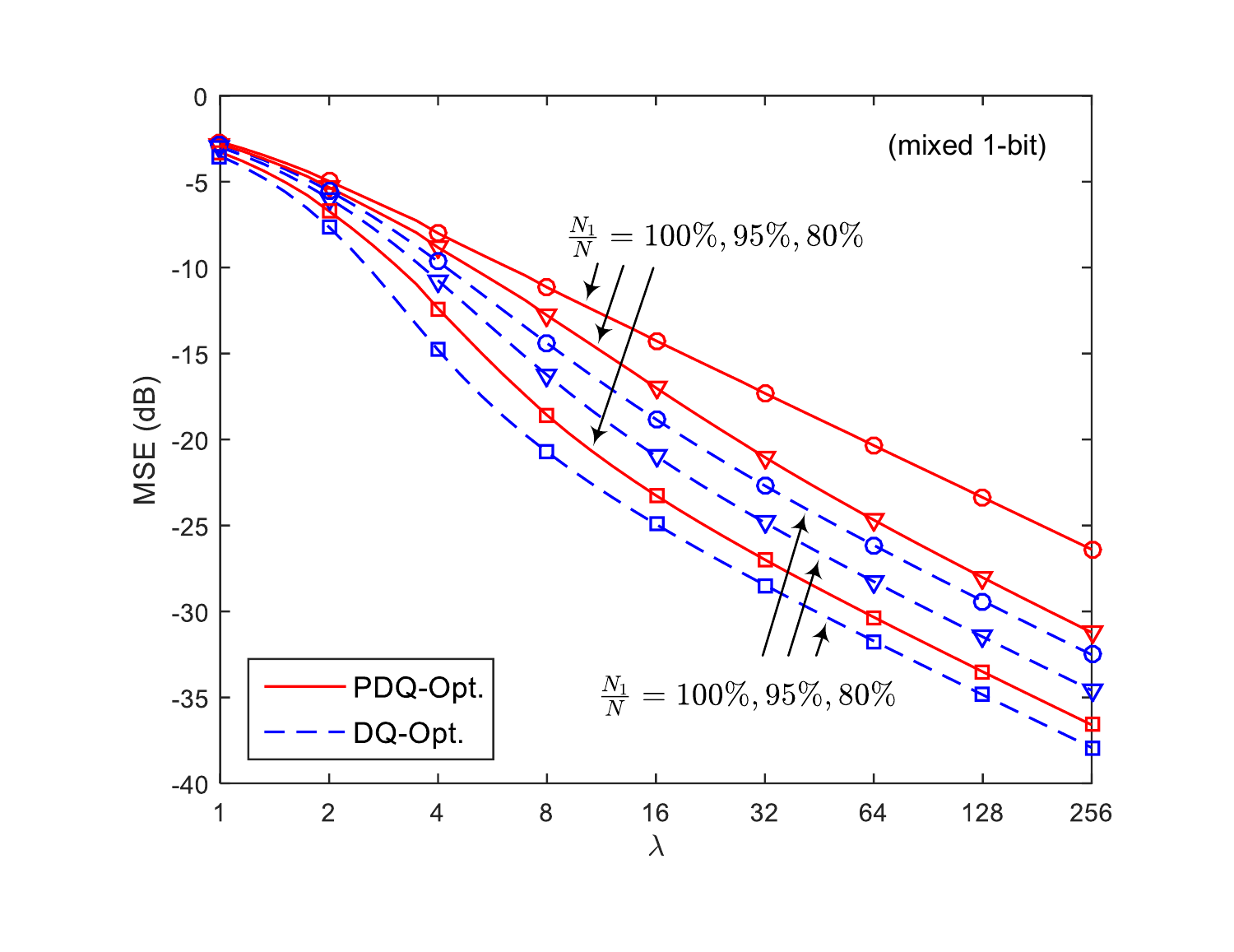} }%
        
        \caption{MSE for a Gaussian input signal versus MIMO configuration $\lambda=N/K$ in a mixed architecture with SNR=$20$\,dB. }\label{fig8}
    \end{center}
\end{figure}

In contrast to Figure~\ref{fig7}, which focuses on the non-mixed architecture, Figure~\ref{fig8} illustrates the results under the mixed architecture. It can be seen that the mixed architecture can help narrow the performance gap resulting from the PQN model. For example, in a massive MIMO system with ${\lambda>10}$, the pure 1-bit PDQ-optimal detector incurs approximately 6\,dB loss compared with the pure 1-bit DQ-optimal detector. Their corresponding gaps are respectively reduced to 3\,dB and 1\,dB when $5\%$ and $20\%$ full precision ADCs are installed.

\section{Conclusion}\label{sec 6}
Using a unified framework, we specified three kinds of detectors for a mixed-ADC massive MIMO receiver by postulating mismatched measures in the Bayes detector. The asymptotic performances of these detectors were analyzed by the SE equations and their accuracy were validated by  Monte Carlo simulations. The SE equations can be quickly and efficiently evaluated; thus, they are useful in system design optimization. We provided useful observations to aid design optimization. In particular, the results showed that we can reduce the computational burden by treating the complex nonlinear quantization process as a PQN model. Ignoring such a complex quantization process may cause 1-bit performance degradation on the high-SNR regime but does not have a significant effect in the low-SNR regime. Moreover, the
mixed receiver architecture can help narrow the performance gap resulting from the PQN model.

\appendices \numberwithin{equation}{section}
\section{Proof of Proposition 1}\label{Appendix A}
In this appendix, we derive the SE equations of Algorithm \ref{alg:GAMP} in the large-system limit (i.e., $N,K\rightarrow \infty$) by following
\cite{mezghani2010belief}. For conceptual clarity and ease of explanation, we restrict the derivation to a corresponding real-valued system although Proposition 1
is for the complex-valued case. In addition, for brevity, we omit all the iteration index $t$.

We recall from lines 7--8 of Algorithm \ref{alg:GAMP} that the conditional mean and variance of $x_j$ denoted by  $\hat{x}_{j}$ and $\hat{v}_{x_{j}}$,
respectively, are taken w.r.t. the marginal posterior $\scP(x_j|s_j,v_{s_j})$ given in (\ref{eq:Px}), which is determined by $(s_j^{t},v_{s_j}^{t})$. Therefore, to
obtain the SE equations, we have to determine the asymptotic behavior of $s_j$ and $v_{s_j}$.

We begin with the analysis of $v_{s_j}$. Let $\mathbf{h}_i^T$ be the $i$-th row of $\mathbf{h}$ and $ w_i = \mathbf{h}_i^T\mathbf{x}$. We notice that given a fixed
$\mathbf{h}_i^T$,
\begin{subequations}\label{A.2}
    \begin{align}
    {\rm E}{\left\{ \mathbf{h}_i^T\mathbf{x} \right\}} &= \sum_{j}h_{ij}\hat{x}_{j} \triangleq \hat{w}_i, \\
    {\rm Var}{\left\{ \mathbf{h}_i^T\mathbf{x} \right\}} &= \sum_{j}h_{ij}^2 \hat{v}_{x_{j}} \triangleq v_{\hat{w}_i},
    \end{align}
\end{subequations}
where the expectation and variance are taken w.r.t. $\scP(x_j|s_j,v_{s_j})$. According to the CLT in the large system limit, $w_i$  can be regarded as a Gaussian
random variable with mean $\hat{w}_i$ and variance $v_{\hat{w}_i}$. Let
\begin{equation}\label{A.3}
\Xi_i(r_i|\hat{w}_i) = \int \ud w \wtP_{\sf out}(r_i|w) e^{ - \frac{(w-\hat{w}_i)^2}{v_{\hat{w}_i}} },
\end{equation}
and we use $\Xi'(r|w)$ and $\Xi''(r|w)$ to denote the first-order and second-order derivatives of $\Xi(r|w)$ with respect to $w$, respectively. Then, $v_{s_j}$ in
line 5 of Algorithm \ref{alg:GAMP} is expressed as
\begin{equation}\label{A.4}
\frac{1}{v_{s_j}} = \sum_{i} \alpha_{ij}^2 - \sum_i\beta_{ij}
\end{equation}
where
\begin{equation}\label{A.5}
\alpha_{ij}=\frac{\Xi_i'(r_i|\hat{w}_{i})}{\Xi_i(r_i|\hat{w}_{i})}h_{ij}
\quad
{\rm and}
\quad
\beta_{ij}=\frac{\Xi_i''(r_i|\hat{w}_{i})}{\Xi_i(r_i|\hat{w}_{i})}h_{ij}^2.
\end{equation}
Letting
\begin{equation}\label{A.6}
A_{j} =\sum_{i}\alpha_{ij}^2 ~~\mbox{and}~~
B_{j} =\sum_{i}\beta_{ij}
\end{equation}
yields
\begin{equation} \label{A.7}
\frac{1}{v_{s_j}} = A_{j} - B_{j}.
\end{equation}
Then, we explore the asymptotic (or the large system) behavior of $A_{j}$ and $B_{j}$.

To this end, we find that when $\mathbf{H}$ is fixed,
\begin{subequations}\label{A.8}
    \begin{align}
    A_{j} & \rightarrow \sum_{i}{\rm E}_{\mathbf{r}}{\left\{\alpha_{ij}^2\Big|\mathbf{H}\right\}} \triangleq \sum_{\kappa}  {\sf A}_{\kappa} , \\
    B_{j}& \rightarrow \sum_{i}{\rm E}_{\mathbf{r}}{\left\{\beta_{ij}\Big|\mathbf{H}\right\}} \triangleq \sum_{\kappa}  {\sf B}_{\kappa},
    \end{align}
\end{subequations}
where $\rightarrow$ indicates the convergence to the asymptotic limit, and we define
\begin{subequations} \label{A.9}
    \begin{align}
    {\sf A}_{\kappa} &= \sum_{i\in \Omega_\kappa}{\rm E}_{\mathbf{r}_\kappa}{\left\{\alpha_{ij}^2\Big|\mathbf{H}\right\}}, \\
    {\sf B}_{\kappa} &= \sum_{i\in \Omega_{\kappa}}{\rm E}_{\mathbf{r}_{\kappa}}{\left\{\beta_{ij}\Big|\mathbf{H}\right\}}.
    \end{align}
\end{subequations}
Here, $\mathbf{r}_{\kappa}$ represents the quantized signals that belong to discrete set ${\cal R}_\kappa$. To compute the expectations in (\ref{A.9}), we need the
joint distribution $P(\mathbf{r}_\kappa,\hat{w}_{i}|\mathbf{H})$ because $\alpha_{ij}$ and $\beta_{ij}$ depend on two correlated variables $\mathbf{r}_\kappa$ and
$\hat{w}_{i}$.

Our strategy to obtain $P(\mathbf{r}_\kappa,\hat{w}_{i}|\mathbf{H})$ is via the marginal of $P(\mathbf{r}_\kappa,\hat{w}_{i},w_{i}|\mathbf{H}) =
P(\mathbf{r}_\kappa,\hat{w}_{i}|w_{i},\mathbf{H})P(w_i,\hat{w}_{i})$. Therefore, we calculate the joint distribution $P(w_i,\hat{w}_{i})$ first. Both $w_i$ and
$\hat{w}_{i}^t$ are sums over many independent terms. Therefore, according to the CLT, they are Gaussian random variables. Their means are zero because
$\{h_{ij}\}$ has zero mean. The entries of the covariance matrix between $w_i$ and $\hat{w}_{i}$ is
\begin{subequations}\label{A.10}
    \begin{align}
    {\rm E}\left\{w_{i}^2|\mathbf{H}\right\}
    &={\rm E}{\left\{\Bigg(   \sum_{j}h_{ij}x_{j}\Bigg)^2\Bigg|\mathbf{H} \right\}} \nonumber \\
    &=\sum_{j}h_{ij}^2{\rm E}{\left\{x_{j}^2|\mathbf{H}\right\}} \nonumber \\
    &\rightarrow \int x^2P_{\rm in}(x)\ud x  \triangleq v_{x}, \\
    {\rm E}\left\{\hat{w}_{i}^2|\mathbf{H}\right\}
    &={\rm E}{\left\{ \left. \Bigg(  \sum_{j}h_{ij}\hat{x}_{ij} \Bigg)^2\right|\mathbf{H}\right\}} \nonumber \\
    &=\sum_{j}h_{ij}^2{\rm E}  \left\{ \hat{x}_{ij}^2 | \mathbf{H} \right\} \nonumber \\
    &\rightarrow \left(\int \hat{x}\scP(\hat{x}) \ud \hat{x}\right)^2 \triangleq v_{\hat{x}}, \label{A.16-2}\\
    {\rm E}\left\{w_{i} \hat{w}_{i} |\mathbf{H}\right\}
    &=\sum_{j}h_{ij}^2  {\rm E}\left\{ \left. x_{j}\hat{x}_{j} \right|\mathbf{H}\right\} \nonumber \\
    &\rightarrow {\rm E}\left\{x  \left(\int \hat{x}\scP(\hat{x}) \ud \hat{x}\right) \right\}
    \triangleq v_{x\hat{x}}.
    \end{align}
\end{subequations}
Here and hereafter, we denote $\scP(x)=\scP(x|s_j,v_{s_j})$ for notation simplicity. We find that the covariance matrix becomes asymptotically independent of the
index $i$. Altogether, these provide the bivariate Gaussian distribution:
\begin{equation}\label{A.11}
P_{w,\hat{w}}(w,\hat{w})
=\mathcal{N}(\hat{w}|0,v_{\hat{x}})\mathcal{N}\left(w\left|\frac{v_{x\hat{x}}}{v_{\hat{x}}}\hat{w},v_x-\frac{v_{x\hat{x}}^2}{v_{\hat{x}}}\right.\right).
\end{equation}

Now, we are ready to calculate the joint distribution $P(\mathbf{r}_{\kappa},\hat{w}_{i}|\mathbf{H})$. For notation simplicity, let
$P(\mathbf{r}_{\kappa},\hat{w}_{i}) = P(\mathbf{r}_{\kappa},\hat{w}_{i}|\mathbf{H})$. Using (\ref{A.11}), we can further calculate
$P(\mathbf{r}_{\kappa},\hat{w}_{i})$ as\footnote{In the derivation, we use the fact that the product of two Gaussians provides another Gaussian, i.e.,
$\mathcal{N}(x|a,A)\mathcal{N}(x|b,B)=Z\mathcal{N}(x|c,C)$, where $c=C(A^{-1}a+B^{-1}b)$, $C=(A^{-1}+B^{-1})^{-1}$, and
$Z=\frac{1}{\sqrt{2\pi(A+B)}}\exp\left(-\frac{(a-b)^2}{2(A+B)}\right)$.}
\begin{align}\label{A.12}
& P(\mathbf{r}_{\kappa},\hat{w}_{i})=\int \ud w_{i}P(\mathbf{r}_{\kappa},\hat{w}_{i},w_{i})\nonumber \\
&=
P(\mathbf{r}_{\kappa \setminus i}) \int \ud w_{i}P( r_{i},\hat{w}_{i},w_{i})\nonumber \\
&=P(\mathbf{r}_{\kappa \setminus i}) \int \ud w_{i}P_{\sf out}(r_{i}|w_{i}) P(w_{i},\hat{w}_{i}) \nonumber \\
&= P(\mathbf{r}_{\kappa \setminus i}) \mathcal{N}(\hat{w}_i|0,v_{\hat{x}})  \nonumber \\
&
\quad \times \int\ud w_i
\mathcal{N}{\left(w_i\left|\frac{v_{x\hat{x}}}{v_{\hat{x}}}\hat{w}_i,v_x-\frac{v_{x\hat{x}}}{v_{\hat{x}}}\right.\right)}
\int_{r_i^{\rm low}}^{r_i^{\rm up}} \ud y \mathcal{N}(y|w_i,\sigma_n^2)
\nonumber \\
&\rightarrow P(\mathbf{r}_{\kappa \setminus i} )\mathcal{N}(\hat{w}_i|0,v_{\hat{x}})
\Psi_i\left(r_i\Bigg|\frac{v_{x\hat{x}}}{v_{\hat{x}}}\hat{w}_i\right),
\end{align}
where $\mathbf{r}_{\kappa \setminus i}$ is the vector containing the elements of $\mathbf{r}_\kappa$ excluding $r_i$, and we define
\begin{equation}\label{A.13}
\Psi_i(r_i|x) = \Phi{\left(\frac{r_i^{\rm up}-x}{\sigma_n^2+v_x-\frac{v_{x\hat{x}}^2}{v_{\hat{x}}}}\right)}
-\Phi{\left(\frac{r_i^{\rm low}-x}{\sigma_n^2+v_x-\frac{v_{x\hat{x}}^2}{v_{\hat{x}}}}\right)}.
\end{equation}

Using the joint distribution (\ref{A.12}), we can compute the asymptotic behavior of (\ref{A.9}) as
\begin{align}\label{A.14}
{\sf A}_{\kappa}&= \sum_{i\in \Omega_{\kappa}}{\rm E}_{\mathbf{r}_{\kappa}}{\left\{\alpha_{ij}^2|\mathbf{H}\right\}} \nonumber\\
&= \sum_{i\in \Omega_{\kappa}}\int \int {\ud\hat{w}_i\ud \mathbf{r}_{\kappa} P(\mathbf{r}_{\kappa},\hat{w}_i){\left(\frac{\Xi'_{i}(y_i|\hat{w}_i)}{\Xi_{i}(y_i|\hat{w}_i)}h_{ij}\right)}^2} \nonumber \\
&=
\sum_{i\in \Omega_{\kappa}} \sum_{r_i\in \mathcal{R}_{\kappa}}\int\ud \hat{w}_i \mathcal{N}(\hat{w}_i|0,v_{\hat{x}})\nonumber \\
& \hspace{0.5cm} \times \Psi{\left(r_i\left|\frac{v_{x\hat{x}}}{v_{\hat{x}}}\hat{w}_i\right.\right)}
{\left(\frac{\Xi'_{i}(r_i|\hat{w}_i)}{\Xi_{i}(r_i|\hat{w}_i)}h_{ij}\right)}^2\nonumber \\
&\rightarrow
\lambda_{\kappa} \sum_{r\in\mathcal{R}_{\kappa}}\int \ud \hat{w} \mathcal{N}(\hat{w}|0,v_{\hat{x}})\Psi{\left(r\left|\frac{v_{x\hat{x}}}{v_{\hat{x}}}\hat{w}\right.\right)}
{\left(\frac{\Xi'(r|\hat{w})}{\Xi(r|\hat{w})}\right)}^2,
\end{align}
and, similarly,
\begin{align}\label{A.15}
{\sf B}_{\kappa}
&\rightarrow
\lambda_{\kappa} \sum_{r\in\mathcal{R}_{\kappa}}\int \ud \hat{w} \mathcal{N}(\hat{w}|0,v_{\hat{x}})
\Psi{\left(r\left|\frac{v_{x\hat{x}}}{v_{\hat{x}}}\hat{w}\right.\right)}\frac{\Xi''(r|\hat{w})}{\Xi(r|\hat{w})}.
\end{align}
Note that ${\sf A}_{\kappa}$ and ${\sf B}_{\kappa}$ are asymptotically independent on the indexes $i,j$ in the large system limit. We also drop the index $i$ of
$\Psi_i(r_i|x)$ defined in (\ref{A.13}). Moreover, we notice that $\Xi_i(r_i|\hat{w}_i)$ defined in (\ref{A.3}) become independent of the index $i$, and read as
\begin{equation}\label{A.15-1}
\Xi(r|\hat{w}) = \int \ud w \wtP_{\sf out}(r|w) e^{ - \frac{(w-\hat{w})^2}{v_{\hat{w}}} },
\end{equation}
where $\hat{w}$ and $v_{\hat{w}}$ are the asymptotic limits of $\hat{w}_i$ and $v_{\hat{w}_i}$, respectively, and $v_{\hat{w}}$  can be expressed as
\begin{align}\label{A.15-2}
v_{\hat{w}}
&= {\rm E}\left\{\sum_{j}h_{ij}^2 \hat{v}_{x_{j}}\Bigg|\mathbf{H}\right\}\nonumber \\
&\rightarrow \int \hat{x}^2\scP(\hat{x})\ud \hat{x}-\left(\int \hat{x}\scP(\hat{x}) \ud \hat{x}\right)^2 \nonumber\\
&\triangleq c_{\hat{x}}-v_{\hat{x}}.
\end{align}
Together with these results, (\ref{A.14}) and (\ref{A.15}) can be further simplified as
\begin{subequations}\label{A.16}
    \begin{align}
    & {\sf A}_{\kappa}
    =\lambda_{\kappa}\sum_{r\in \mathcal{R}_{\kappa}}
    \int\uD u \Psi{\left(r\left|\sqrt{\frac{v_{x\hat{x}}^2}{v_{\hat{x}}}}u\right.\right)}
    {\left(\frac{\Xi'(r|\sqrt{v_{\hat{x}}}u)}{\Xi(r|\sqrt{v_{\hat{x}}}u)}\right)}^2, \\
    & {\sf B}_{\kappa}
    =\lambda_{\kappa}\sum_{r\in \mathcal{R}_{\kappa}}
    \int\uD u \Psi{\left(r\left|\sqrt{\frac{v_{x\hat{x}}^2}{v_{\hat{x}}}}u\right.\right)}\frac{\Xi''(r|\sqrt{v_{\hat{x}}}u)}{\Xi(r|\sqrt{v_{\hat{x}}}u)}.
    \end{align}
\end{subequations}
Thus, we obtain the asymptotic behavior of $1/v_{s_j}$ in (\ref{A.7}) as
\begin{equation} \label{A.17}
\frac{1}{v_{s_j}} \rightarrow A - B \triangleq E,
\end{equation}
where $A = \sum_{\kappa} {\sf A}_{\kappa}$ and $B = \sum_{\kappa} {\sf B}_{\kappa}$ with ${\sf A}_{\kappa}$ and ${\sf B}_{\kappa}$ given in (\ref{A.16}). Note that $1/v_{s_j}$ becomes
asymptotically independent of the index $j$.

Next, we consider the asymptotic behavior of $s_{j}$ given in line 6 of Algorithm \ref{alg:GAMP}. Following the similar argument presented in this paper, we can
prove that $s_j$ conditioned on $x_j$ is asymptotic Gaussian. Again, we drop the index $j$ in the large system limit, and we show that the asymptotic mean and
variance of $s$ is $\frac{D}{E} \cdot x$ and $\frac{A}{E^2}$ with $D = \sum_{\kappa} {\sf D}_{\kappa}$, where
\begin{equation}\label{A.18}
{\sf D}_{\kappa}=\lambda_{\kappa}\sum_{r\in \mathcal{R}_{\kappa}}
\int\uD u \Psi'\left(r\Bigg|\sqrt{\frac{v_{x\hat{x}}^2}{v_{\hat{x}}}}u\right)\frac{\Xi'(r|\sqrt{v_{\hat{x}}}u)}{\Xi(r|\sqrt{v_{\hat{x}}}u)},
\end{equation}
and $\Psi'(r|u)$ represent the derivatives of $\Psi$ with respect to $u$. Therefore we denote
\begin{equation}\label{A.19}
s = \frac{D}{E} x+\sqrt{\frac{A}{E^2}}z,
\end{equation}
where ${z\sim\mathcal{N}(0,1)}$ and $x \sim P_{\sf in}(x)$. Using (\ref{A.17}) and (\ref{A.19}), the marginal posterior (\ref{eq:Px}) can be rewritten as
\begin{align}\label{A.20}
&\scP(\hat{x}|x,z) = \frac{ \wtP_{\sf in}(\hat{x})\exp{\left(-\frac{\left(\hat{x}-\frac{D\cdot x+\sqrt{A}z}{E}\right)^2}{2\frac{1}{E}}\right)}}{\int  \wtP_{\sf in}(\hat{x})\exp{\left(-\frac{\left( \hat{x}-\frac{D\cdot x+\sqrt{A}z}{E}\right)^2}{2\frac{1}{E}}\right)\ud \hat{x}    }}.
\end{align}

Substituting (\ref{A.20}) into (\ref{A.10}) and (\ref{A.15-2}), the parameters $v_x$, $c_x$, $v_{\hat{x}}$, and $v_{x\hat{x}}$ can be obtained, which are
\begin{subequations}\label{A.21}
    \begin{align}
    v_x&=\mathbb{E}_x\left\{x^2\right\}, \\
    c_{\hat{x}}&={\rm E}_{x,z}{\left\{\int\hat{x}^2\scP(\hat{x}|x,z)\ud \hat{x}\right\}},     \\
    v_{\hat{x}}&={\rm E}_{x,z}{\left\{\left(\int \hat{x} \scP(\hat{x}|x,z)\ud \hat{x} \right)^2\right\}}, \\
    v_{x\hat{x}}&={\rm E}_{x,z}{\left\{x\int \hat{x} \scP(\hat{x}|x,z)\ud \hat{x}\right\}}.
    \end{align}
\end{subequations}

Finally, we conclude that the asymptotic behavior of $s_j$ in Algorithm \ref{alg:GAMP} can be characterized by (\ref{A.19}), where the parameters $A,D,E$ are
determined by (\ref{A.16}) and (\ref{A.18}) in conjunction with the parameters $v_x,c_x,v_{\hat{x}},v_{x\hat{x}}$ given in (\ref{A.21}). For the complex-valued
case, the signal power of the real or imaginary part is $1/\sqrt{2}$. Therefore, the equivalent quantization interval should multiply a factor of $\sqrt{2}$ as in
(\ref{4.3}). Besides, the relative scalar multiplication should be modified to complex multiplication as in (\ref{4.2}) and (\ref{4.4}).

\end{document}